\documentclass[12pt]{article}

\usepackage{scicite}

\usepackage[T1]{fontenc}
\usepackage{textcomp}
\usepackage{gensymb}
\usepackage{amsmath} 
\usepackage{color}
\usepackage{graphicx}
\usepackage{xfrac}
\usepackage{soul} 
\usepackage{lineno} 
\usepackage[english]{babel}
\usepackage{blindtext}
\usepackage{amssymb}
\usepackage{caption}
\usepackage{array}
\usepackage{makecell}
\usepackage{comment}
\usepackage{float}
\usepackage{hyperref}
\usepackage{url}
\usepackage{siunitx}

\usepackage{times}

\newenvironment{sciabstract}{%
\begin{quote} \bf}
{\end{quote}}

\renewcommand\refname{References and Notes}

\newcounter{lastnote}
\newenvironment{scilastnote}{%
\setcounter{lastnote}{\value{enumiv}}%
\addtocounter{lastnote}{+1}%
\begin{list}%
{\arabic{lastnote}.}
{\setlength{\leftmargin}{.22in}}
{\setlength{\labelsep}{.5em}}}
{\end{list}}

\hypersetup{
    colorlinks=false,
    linkcolor=blue,
    filecolor=magenta,      
    urlcolor=cyan,
    pdftitle={Overleaf Example},
    pdfpagemode=FullScreen,
    }

\graphicspath{ {./Figures/} }

\makeatletter
\DeclareRobustCommand\bfseriesitshape{%
  \not@math@alphabet\itshapebfseries\relax
  \fontseries\bfdefault
  \fontshape\itdefault
  \selectfont
}
\makeatother

\newcommand{\beginsupplement}{%
        \setcounter{table}{0}
        \renewcommand{\thetable}{S\arabic{table}}%
        \setcounter{figure}{0}
        \renewcommand{\thefigure}{S\arabic{figure}}%
     }

\title{PASS: An Asynchronous Probabilistic Processor for Next Generation Intelligence}

\author{Saavan Patel$^{1, 2 *}$, Philip Canoza$^1$, Adhiraj Datar$^1$, Steven Lu$^1$, \\ Chirag Garg$^1$, Sayeef Salahuddin$^{1, *}$\\
\\
\normalsize{${}^{1}$Department of Electrical Engineering and Computer Sciences}\\ 
\normalsize{University of California, Berkeley, California 94720, USA } \\ \\
\normalsize{${}^{2}$Present Address: InfinityQ Technology Inc} \\ \normalsize{Montreal, Quebec H3C 1E2, Canada}\\
\\
\normalsize{$^\ast$To whom correspondence should be addressed;} \\
\normalsize{E-mail:  saavan@berkeley.edu, sayeef@berkeley.edu}
}

\date{}

\begin{document}

\maketitle


\begin{sciabstract}

New computing paradigms are required to solve the most challenging computational problems where no exact polynomial time solution exists.Probabilistic Ising Accelerators has gained promise on these problems with the ability to  model complex probability distributions and find ground states of intractable problems. In this context, we have demonstrated the Parallel Asynchronous Stochastic Sampler (PASS), the first fully on-chip integrated, asynchronous, probabilistic accelerator that takes advantage of the intrinsic fine-grained parallelism of the Ising Model and built in state of the art 14nm CMOS FinFET technology. We have demonstrated broad applicability of this accelerator on problems ranging from Combinatorial Optimization, Neural Simulation, to Machine Learning along with up to $23,000$x energy to solution improvement compared to CPUs on probabilistic problems.

\end{sciabstract}

\section*{Introduction}

 Accelerators based on the Ising model have gained interest in the recent years due to their close connection to a wide variety of fields, including Machine Learning \cite{Ackley1985AMachines, Hinton2002TrainingDivergence, Onizawa2020In-HardwareLearning}, Neurobiology, \cite{Sridhar2021TheCollectives, Hopfield1982NeuralAbilities.}, Combinatorial Optimization \cite{Lucas2014IsingProblems, Patel2022LogicallyFactorization}, Quantum physics \cite{Camsari2019Scalable-bits} and much more. Ising Model accelerators can be divided into two broad classes: (i) annealing based methods that aim to find the ground state of a specific input problem, and (ii) probabilistic methods which can fully model the solution space of a given problem and are more broadly applicable due to their ability to model distributional quantities, such as variable expectation, uncertainty and distributional distances \cite{Ma2019SamplingOptimization, Goodfellow2016DeepLearning}. Although there have been many implementations of annealers \cite{Moy2022ASolving, Blais2000OperationSuperconductors}, implementations of probabilistic computing have remained elusive. The neurons in a probabilistic computer (p-bits) needs to provide a stochastic output whose statistics should depend on the simultaneous input of the input neighbor p-bits. Additionally, to take advantage of the parallelism inherent in such an architecture, interaction between the neurons needs to be asynchronous. Both of these aspects pose substantial challenge for physical realization which has meant that existing demonstrations of probabilistic computing systems (p-bits) \cite{Camsari2017ImplementingMTJ, Borders2019IntegerJunctions, Hayakawa2021NanosecondJunctions, Shao2023ProbabilisticJunctions} have mostly relied on single device demonstrations as well as emulations of asynchronous operation on a digital hardware such as FPGA. \cite{Hayakawa2021NanosecondJunctions, Shao2023ProbabilisticJunctions, Patel2020IsingMachine, Patel2022LogicallyFactorization}. In this work, we have gone beyond these conventional approaches by demonstrating a fully integrated, asynchronous, probabilistic computer, capable of parallel sampling of the solution states. In the interest of brevity, we will call this hardware the PASS (Parallel Asynchronous Stochastic Sampler) accelerator.

 In its heart, the PASS is based on the Boltzmann Machine \cite{Ackley1985AMachines} binary neural network (shown below in equation \ref{eq:boltzmann}) that mimics how the brain computes in a massively parallel, distributed, and asynchronous manner. PASS follows a sampling-based approach in which each neuron updates asynchronously based on its neighbors following a Poisson clock generated by inherent shot-noise present in advanced CMOS nodes. This method of computing offers a theoretical scaling advantage over traditional digital computing \cite{Hayes2007AGraphs}, where neurons are updated synchronously and in order. This method also expands the flexibility to mapping of more arbitrary graphical models by not being restricted to the fixed update scheme or connectivity. While the connectivity scheme in this iteration of PASS is fixed to nearest neighbors and diagonals, future iterations can have next nearest neighbor and higher order interactions without any change to the underlying neuron or connection fabric architecture, but by simply increasing the size of the digital binary dot product component. 
\begin{gather}
\label{eq:boltzmann}
    p(s) = \frac{1}{Z} e^{-E(s)} \\ 
    E(s) = \sum_i \sum_j J_{ij} s_i s_j + \sum_i b_i s_i, \ \  s \in \{-1, 1\}
\end{gather}

Central to the PASS accelerator are three salient features: (i) A stochastic neuron that takes advantage of the shot noise present in electrical circuits. Notably, suppressing noise is of paramount importance in conventional circuits; instead, we have leveraged noise to achieve a stochastic neuron that offers an exceptionally fast stochastic response ($\approx\si{ns}$) compared to existing implementations  \cite{Borders2019IntegerJunctions, Hayakawa2021NanosecondJunctions, Shao2023ProbabilisticJunctions}  (ii) An in-memory computing architecture with a flexible digital fabric that connects the neurons through a fully asynchronous neural network with 8 bit weights and biases (iii) Ability to program and integrate this system with traditional computers that allows us to map and solve a broad class of problems.

PASS also takes advantage of full-stack (algorithm-architecture-technology) design. Firstly, the use of a sampling-based architecture allows for an efficient hardware design. PASS uses a low precision, fixed point, weight-stationary architecture, causing all calculations to be extremely hardware efficient, reducing the area, power and time to solution. 

Secondly, noise that helps the system escape local minima is inherent to the sampling based scheme of the PASS accelerator. The ad-hoc integration of noise, which many traditional accelerators rely on, limits how closely the hardware resembles the algorithm which they aim to adopt \cite{Yamaoka2016AAnnealing, Wang2013CoherentOscillators, Lo2023AnArchitecture}. From a theoretical perspective, the PASS accelerator is designed to closely match the stochastic Glauber Dynamics \cite{Glauber1963TimeDependentModel} In fact, recent work has shown that sampling-based algorithms can outperform annealing-based algorithms on certain optimization tasks \cite{Ma2019SamplingOptimization}. The probabilistic form of interaction means that while many other systems will use noise simply to escape local minima, the PASS system uses noise to stochastically model the full probability distribution underlined by the inputs. This means that all of the intermediate states seen in the PASS system have a meaning and allow for probabilistic calculations, like calculation of expectation values (useful in machine learning experiments), identification of local attractors (useful in neural simulation), and identifying important parameters of the distribution like the mode and local minima (useful in optimization). This allows for a more flexible form of computation which can be useful as a general purpose computing device. 

Lastly, PASS relies on a fundamentally localized form of interaction, creating a system which is fully parallel in its computation, with each neuron updating simultaneously. This architecture is accomplished by using asynchronous updates, where each neuron probabilistically updates its state whenever the value of its neighbors changes. When comparing a synchronous accelerator operating at the same effective clock speed as PASS, we observed a $\sim 200$x time to solution increase by using the PASS scheme (Fig \ref{fig:maxcut}).  While a factor of $k$ improvement (where $k$ is the number of connections per neuron) is theoretically derived \cite{Hayes2007AGraphs},in this work we see a greater than linear improvement in MaxCut and SK problems. Further details of the simulation model are discussed in the Methods section (in the Supporting Materials). This type of update scheme mimics the way the brain computes and demonstrates a neuromorphic approach to computing within the Ising Model \cite{Hopfield1982NeuralAbilities., Ackley1985AMachines}.  

To demonstrate the broad applicability of the PASS accelerator, we have applied it to problems in three spaces; (i) optimization problems where we demonstrated how this system could improve upon current state of the art methods, (ii) Machine Learning where we showed how these physical samplers could be used to do multiplier-free machine learning, eliminating the most expensive computational component in modern machine learning systems and (iii) Neural simulation where we demonstrated successful modeling of real-time decision making in primitive animal brains.

\section*{Hardware Design}

As mentioned earlier, we used the inherent shot noise present in advanced CMOS nodes to create a binary, stochastic neuron circuit. The neurons were coupled with adjacent neurons through a connection circuit comprising of a digital fabric in an analog-mixed signal architecture (Fig \ref{fig:hardware} A). The neurons were composed of two pieces, a synapse and a stochastic neuron circuit. The synapse design was composed of a dot product engine, which takes the binary activations of adjacent neurons to mask and accumulate them, and a Digital to Analog Converter (DAC) which passed the accumulated values into the neuron. The interaction strength with adjacent neurons was stored as an 8-bit fixed point number to allow for complex interaction representations. The neuron circuit had three major components; a noise source and noise amplifier which generated a stochastic signal for every neuron, a sigmoidal activation function which combined the stochastic signal from the neuron with the output of the synapse circuit, and the output buffer which digitized the output and drove the signal to other parts of the chip. The extracted activation function from a PASS chip is shown in Fig \ref{fig:hardware} B, along with time series signals Fig \ref{fig:hardware} C, D, and E which show the output voltage with respect to time at 3 different input voltages to the neuron. From these time series signals we showed that the stochastic voltage fluctuates on $\mu$-second timescales between 0 and Vdd, and the output average voltage was modulated effectively by the input voltage. The PASS chip at various size scales is demonstrated in Fig \ref{fig:hardware} F), G) and H) where we saw the single neuron post-layout diagram, as well as 4 interconnected neurons and finally the full chip post-layout including neuron core, I/O, and sampling circuitry. Further detail on the hardware design is explained in the methods section. 

More arbitrary graphical models can be accomplished by increasing the degree of connectivity of each neuron (such as incorporating next nearest neighbor interactions), and by incorporating higher order interactions by increasing the complexity of the dot product engine. The lack of fixed update scheme means that as long as the increase in neuron connectivity does not increase synapse delay significantly (we expect increase in connectivity to create a $log(k)$ increase in synapse delay through the binary adder tree section \cite{Hennessy2012ComputerApproach}) then we can increase the number of connections. Additionally, the decentralized spatial compute fabric allows for the system to scale up depending on silicon area without arising from communication delays due to clocks and similar centralized systems. Overall this means that the PASS system allows for a large degree of connectivity that is not available in many other systems. 

To be able to integrate a full system on-chip with with a high degree of connectivity, a number of hardware problems were tackled. Most notably, neuron to neuron variation, neuron communication delay, and high precision connectivity were taken into account to be able to solve large problems. Neuron connectivity and high precision were effectively tackled by using a mixed-signal architecture where the digital components of the circuit handle accumulations and weights memory. While this adds complexity, area and power, due to the need for a DAC within each neuron, the additional flexibility was crucial to tackle problems requiring high precision (See discussion of architecture in Methods section). Neuron to neuron variation was tackled through the use of analog trim circuitry for each neuron, post-fabrication correction to individual chip (see Supplementary Figure \ref{fig:sigmoid_correction}), and under-voltage of the chip (See Supplementary Figure \ref{fig:voltage_variation} and discussion in Methods). Neuron communication delay was analyzed and characterized in both simulation and real hardware, and can be adjusted using the analog trim circuitry present in each chip (See Supplementary Figure \ref{fig:delay_hw}, \ref{fig:delay_sim}, and discussion in Methods). 
 
\section*{Optimization}

Ising solvers have found many applications in the acceleration of the solution of combinatorial optimization problems. Accordingly, we addressed combinatorial optimization as our first demonstrative example. Fig \ref{fig:maxcut} A shows a MaxCUT problem of 4 nodes, where the system correctly identified the two states that correspond to the ground state of the problem. This `toy-level' exercise allowed us probe the exact voltage and time dynamics of the fluctuating system, as demonstrated in Fig \ref{fig:maxcut} B, C, D and E.  We observed that the system fluctuates between the two ground state solutions on $\mu$-second timescales. Next, we constructed an artificial MaxCUT problem that involves all the neurons in our system. Here the ground state was analytically defined to spell out the letters C, A and L. Such a MaxCUT problem was constructed by having ferromagnetic (positive) coupling between regions of the same color and having opposing coupling in the transition regions between the letter and the background. PASS accelerator successfully reached the ground state spelling the correct letters (Fig \ref{fig:maxcut} G), which demonstrated the ability of the larger system to reach global minima in an asynchronous fashion.

To demonstrate the algorithmic speed improvements that come with the asynchronous update scheme, we compared the performance of the PASS accelerator to that of a discrete time accelerator with the same effective clock frequency (the same average update time) (see Fig \ref{fig:maxcut} H).  We found that the time to solution shows a $200$x acceleration for the PASS accelerator compared to a discrete time solution on 150 node problems when compared with the same update rate. While theory suggested that the asynchronous update scheme would support a factor of $n$ scaling improvement as the problem size grows, we found an additional empirical advantage in the exponential scaling constant when fits are done on both the MaxCut and SK problems (with p<0.01 the scaling parameters are not the same, see Supplementary Table \ref{tab:scaling_fit_params} and Supplementary Figure \ref{fig:SK_MC_scaling_fit}). While this is a notable empirical result, this would benefit from further exploration from a theoretical perspective in further works (See Methods section for further details). This result underscored the importance of asynchronous operation in achieving high acceleration and potential scaling advantage. Compared to existing state of the art accelerators PASS simulations demonstrated the lowest time to solution for the same benchmark MaxCUT problem (Fig \ref{fig:maxcut} I). We note here, however, that it is quite difficult to do a one-on-one comparison because the exact nature of connectivity, uncertainty about latency connected to assisting technologies, pre-processing of data etc. vary among reported demonstrations. Additionally, while the PASS system shown here does not perform simulated annealing, simulated annealing is possible in future systems by having a counter that uniformly decreases the value of the weights as computation progresses \cite{Geman1987StochasticImages, Bertsimas1993SimulatedAnnealing}. 

\section*{Machine Learning}
 
\begin{figure*}
\begin{centering}
\includegraphics[width=\linewidth]{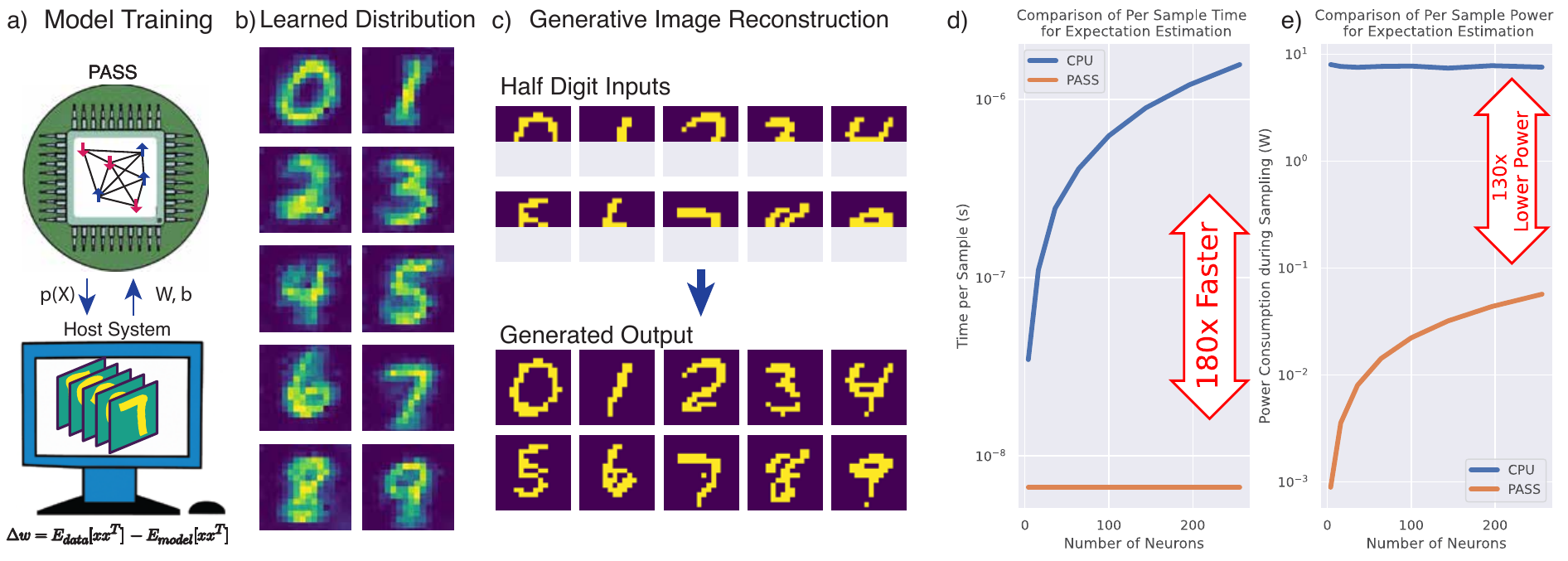}
\par\end{centering}
\caption{\label{fig:ML}\textbf{ Multiplier Free Generative Machine Learning}  \\
\textbf{ (A)\/} Diagram of overall machine learning system with the PASS system. The host system holds the training data, which it uses to calculate the data expectation, and the PASS system computes the model expectation with the given Weights and Biases. The host system then calculates changes in weights based on this and iterates until the model has converged. None of the operations (expectations, binary outer products, averaging) require multiplications due to the binary nature of activations and the PASS stochastic activation system. 
\textbf{ (B)\/} An example of learned digit distributions taken from the MNIST dataset. The PASS system is trained on each digit individually, these images show the average activations after being trained on the given digit.
\textbf{ (C)\/} After learning the digit distributions, the PASS system can perform generative modeling tasks, such as image reconstruction given partial images. The system is clamped with the top half of a digit (top figures), and the bottom represents a sampled output from the system, showing that it can effectively model the given distribution.
\textbf{ (D)\/} PASS is able to produce samples $180$x faster with a flat scaling resulting from the ability to fully utilize the parallelism of the PASS platform. The CPU is running. This yields an extremely power and time efficient platform for machine learning. 
\textbf{ (E)\/} PASS is able to produce samples with a power consumption of $\approx 130x$ during the sampling run using $\approx 130$x less power for full chip simulation (222 \si{\mu W} per neuron  and 56.8 \si{mW} full chip vs. 7W for CPU power consumption on a single core). This yields an overall $23,400$x improvement in energy to solution to produce a given number of samples for a machine learning experiment.}
\end{figure*}

As a second demonstrative example, we explored generative machine learning using the PASS accelerator. Notably, the basis for the PASS accelerator is the Boltzmann Machine framework, a generative model, where the full probability distribution of both input and outputs is used to describe a problem (see Equation \ref{eq:boltzmann}). The PASS accelerator enabled accelerated sampling from complex and correlated distributions.  To use it as a training accelerator, we connected the PASS chip to a classical computer which calculates the gradients for weights ( Fig \ref{fig:ML} A). This configuration was next used to explore the MNIST database \cite{Lecun1998Gradient-basedRecognition} in a generative fashion. We observed that the accelerator is able to learn digit distributions from training data accurately (Fig \ref{fig:ML} B). Next, we showed a generative modeling task, where we input partial images and the PASS accelerator attempts to reconstruct the other half of the image. Fig \ref{fig:ML} C) shows that a successful reconstruction could be achieved. Compared to a traditional CPU running the same algorithm, PASS showed a $180$x speed improvement for generating each sample in the distribution as shown in Fig \ref{fig:ML} D). When examining the power consumption, we see that the accelerator consumed $\approx 130$x less power as demonstrated in Figure \ref{fig:ML} E) . Importantly, there was also a large scaling improvement for the PASS accelerator; the time per sample was nearly constant due to the parallel nature of the updates (Fig \ref{fig:ML} D) as opposed to the the serial nature of CPU. Further details on these experiments are provided in the Methods section.

Notably, the contrastive divergence algorithm simply relies on averages, additions and subtractions shown below in Equation \ref{eq:CD}. Specifically, the $s_is_j$ terms are binary AND gates (as both are binary activations), and the averaging across the expectation can be accomplished simply by bit-shifting and adding. This means that the PASS system can perform both training and inference without using any multiplier blocks. 
\begin{equation}
    \label{eq:CD}
    \Delta w_{ij} = \alpha (\mathbb{E}[s_is_j]_{data} - \mathbb{E}[s_is_j]_{model})
\end{equation}
The majority of compute area and power for many machine learning systems is devoted to multiplying gradients for complicated back-propagation through many deep learning layers during training and for multiplication of forward activations during inference \cite{Jouppi2017In-DatacenterUnit, Jouppi2021}. The energy cost of performing a multiplication is approximately $10$x higher than the cost of doing an addition (for Int8 precision \cite{Jouppi2021}) suggesting that removing the multiplication step would have a large impact on the energy to train and perform inference, validating the large energy savings we are showing here. The removal of the multiplication step, together with the inherent parallelism achieved through asynchronous operation, could lead to substantial performance improvement for Markov Chain Monte Carlo based methods for training \cite{Hinton2002TrainingDivergence}.

\section*{Neural Decision Making}
\begin{figure*}
\begin{centering}
\includegraphics[width=\linewidth]{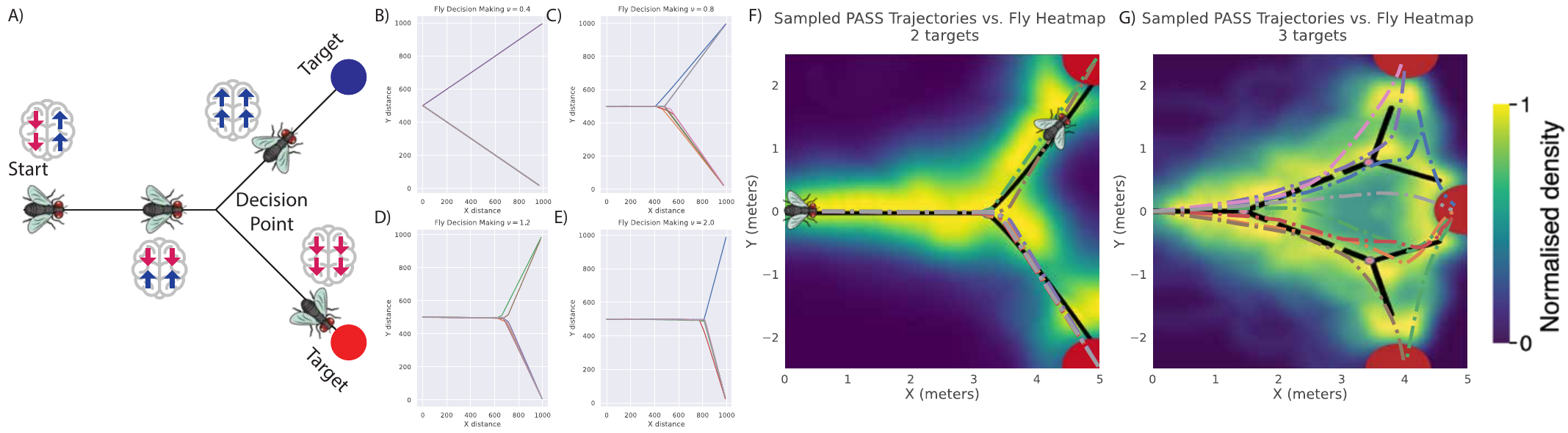}
\par\end{centering}
\caption{\label{fig:fly}\textbf{ Neural Decision Making in primitive brains using the PASS system}  \\
\textbf{ (A)\/} Diagram showing how the Ising model is mapped onto decision making in primitive fly brains. As the fly moves closer to the targets, the neurons spontaneously make a collective decision about which target to approach based on stochastic ring attractor dynamics. 
\textbf{(B)\/},\textbf{(C)\/},\textbf{(D)\/},\textbf{(E)\/} Showing how the neural tuning parameter $\eta$ effects the geometry of the space that the fly operates in. As $\eta$ increases, the fly makes decisions closer to the targets. Targets are placed at \{0, 1000\}  and \{1000, 1000\}. 
\textbf{ (F)\/} When choosing the tuning parameter of $\eta=1.0$ for we see that the sampled trajectories from the PASS chip (the colored dotted lines) match closely with actual trajectories from flies placed into a virtual reality environment. The heatmap shows density for actual fly trajectories placed into a virtual reality environment with two targets. 
\textbf{ (G)\/} The PASS chip sampling trajectories for the 3 target case. Sampled trajectories show random decisions associated with fly trajectories maintaining discrete decision points associated with the targets. }
\end{figure*}

As a third example, we explored the problem of modeling neural decision making. The Ising model and the Hopfield Network have been used as effective tools in neurobiology for many years to model the behavior of firing neurons within the brain \cite{Hopfield1982NeuralAbilities.}. More recently an Ising type model was used to model animal decision making in primitive animal brains \cite{Sridhar2021TheCollectives}. In this system, flies and other primitive animals were placed in a virtual reality environment, and were given a set of targets. It was found that the animals would move to the average of the targets, and spontaneously bifurcate to make a decision about the target they would proceed to. This kind of decision making is graphically shown in Fig \ref{fig:fly} A where the fly chooses between the red and blue target, spontaneously choosing which direction to go at a decision point that is given by the animal's geometric interpretation of the space. The neuron couplings between adjacent spins represent interactions that incorporate the geometry of the space in which the flies reside. The equation of neuron couplings is represented by the equation below, where the pairwise couplings are set by the angle ($\theta_{ij}$) between the goal vectors associated with those spins and $\eta$ is used as a free parameter.
\begin{equation}
    \label{eq:main_fly_coupling}
    J_{ij} = cos(\pi(\frac{|\theta_{ij}|}{\pi})^\eta)
\end{equation}
We successfully mapped this problem onto the PASS accelerator and performed simulations directly on the hardware (see Methods for more details). In these experiments, the PASS accelerator solves Equation \ref{eq:main_fly_coupling} to simulate the fly making the next decision for direction, while calculations are done off-chip to re-calculate the velocity of the fly, the new position, and the new interaction coefficients. The results are shown in Fig \ref{fig:fly} B, C, D, and E. As $\eta$ is increased, the point of spontaneous decision moves later and later in the overall trajectory. This decision point is dependent on the neurobiology of both the individual and the species being explored, demonstrating that the PASS system can model a variety of types of systems. When tuning this $\eta$ parameter, we see that the PASS sampled trajectories in Fig \ref{fig:fly} F (shown as colored dashed lines) match up very similarly to the normalized density of real fly trajectories taken from physical experiments \cite{Sridhar2021TheCollectives}. In more complex situations, like the three target examples shown in Fig \ref{fig:fly} G, we saw that the PASS sampled trajectories still matched up with the densities given by real fly systems and also exhibiting the stochasticity expected from real animal decision making. 

This model of decision making relies on the system probabilistically falling into local attractors reflecting a particular decision, rather than directly finding ground state solutions. In fact, this kind of neural computation is essentially probabilistic, with the circuit noise in the PASS chip mimicking the neural noise present in animal brains \cite{Sridhar2021TheCollectives, Zhang1996RepresentationTheory} which is a crucial piece of understanding neural behavior. Additionally, while experiments shown here are done with symmetric connections (relating to the boltzmann machine framework), asymmetric connections are implemented and possible in this PASS architecture, leading to deeper emulation of the asymmetric connections and non-equillibrium physics of the brain \cite{Zhang1996RepresentationTheory}. The decision making that the PASS system exhibited operated at much faster time scales compared to true fly decision making, with each neuron sampling run taking $\approx 41 \mu s$ compared to fly decision making which occurs over $\approx 100 \si{ms}$ \cite{DasGupta2014FoxPDrosophila}. This suggests that larger scale PASS systems could have the ability to emulate real time decision making in more complex brains, such as those of mammals.

\section*{Conclusion}
Probabilistic, asynchronous solution of Ising models has been discussed extensively in literature \cite{Camsari2017ImplementingMTJ, Aadit2022MassivelyMachines, Aadit2023AcceleratingP-bits, Sutton2020AutonomousSecond, Rhee2023ProbabilisticPbit, Mahmoodi2019VersatileNeurooptimization}. Our demonstration is the first fully integrated system that incorporates both these features. We believe that many of these works can additionally be beneficial to the next generation of the PASS system by combining elements of stochastic dot products \cite{Mahmoodi2019VersatileNeurooptimization} to alleviate computational burdens, new materials for better noise sources \cite{Camsari2017ImplementingMTJ, Rhee2023ProbabilisticPbit}, and improved hardware autonomous operations \cite{Sutton2020AutonomousSecond}. The implementation itself answers a very critical question on the use of noise; it shows that simply using the noise in electrical circuits is sufficient to provide appropriate sample distributions and no  sophisticated digital or analog random number generation is strictly necessary. This together with the asynchronous operation demonstrates that naturally updating neurons, without any clock, can capture the probability distribution function of a complex landscape. The result is high acceleration, as we have shown for multiple different problems. Our demonstration in one of the most advanced CMOS technology also shows that scaling to very large systems is readily possible. There has been extensive work especially in new materials and systems which help to enable higher degrees of connectivity in next generation systems \cite{Lo2023AnArchitecture, Mahmoodi2019VersatileNeurooptimization}. The probabilistic nature allows solution of wide variety of problems beyond what traditional annealers can tackle. We have demonstrated examples of Combinatorial Optimization, Generative Machine Learning and Neural Simulation. But there are many other examples possible such as Quantum Computation \cite{Chowdhury2023AcceleratedComputers, Chowdhury2023EmulatingMachines}, Physics and Material Simulation \cite{Huang2016ASimulations}, and much more. Accelerating time to solution, reducing the power required, and enabling solutions of large and complex problems could substantially benefit edge applications in resource constrained environments.


\newpage



\clearpage
\begin{figure*}

\begin{centering}
\includegraphics[width=\linewidth]{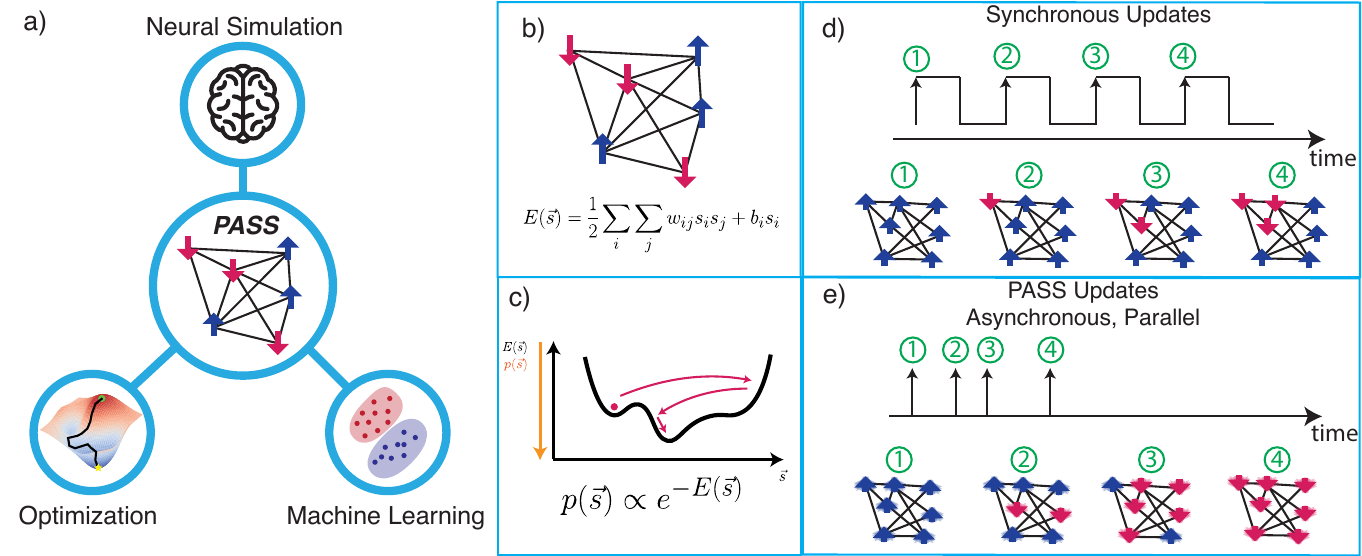}
\par\end{centering}
\caption{\label{fig:alg}. \textbf{Representation of the PASS algorithm and system} \\ 
{\textbf{(A)}} The PASS system relies on a system of interacting spins, and has the ability to map to applications in many areas including Optimization, Machine Learning and Neural Simulation. 
{\textbf{(B)}} We follow a Boltzmann Machine approach, where we are looking to probabilistically sample over a system given by pairwise interacting spins. 
{\textbf{(C)}} Sampling proceeds by finding low energy configurations, which map to high probability states. 
{\textbf{(D)}} The traditional synchronous update scheme for a Boltzmann System would have each update occur on a fixed clock scheme, as shown here. This only allows one variable to update at a time, slowing down the overal update rate. 
{\textbf{(E)}} By following asynchronous update schemes, each neuron is not forced to use a fixed update scheme, and can update continuously based on its neighboring values. This causes a large, embarassingly parallel approach drastically speeding up convergence over the underlying sampling algorithm. 
}

\end{figure*}

\clearpage
\begin{figure*}
\begin{centering}
\includegraphics[width=\linewidth]{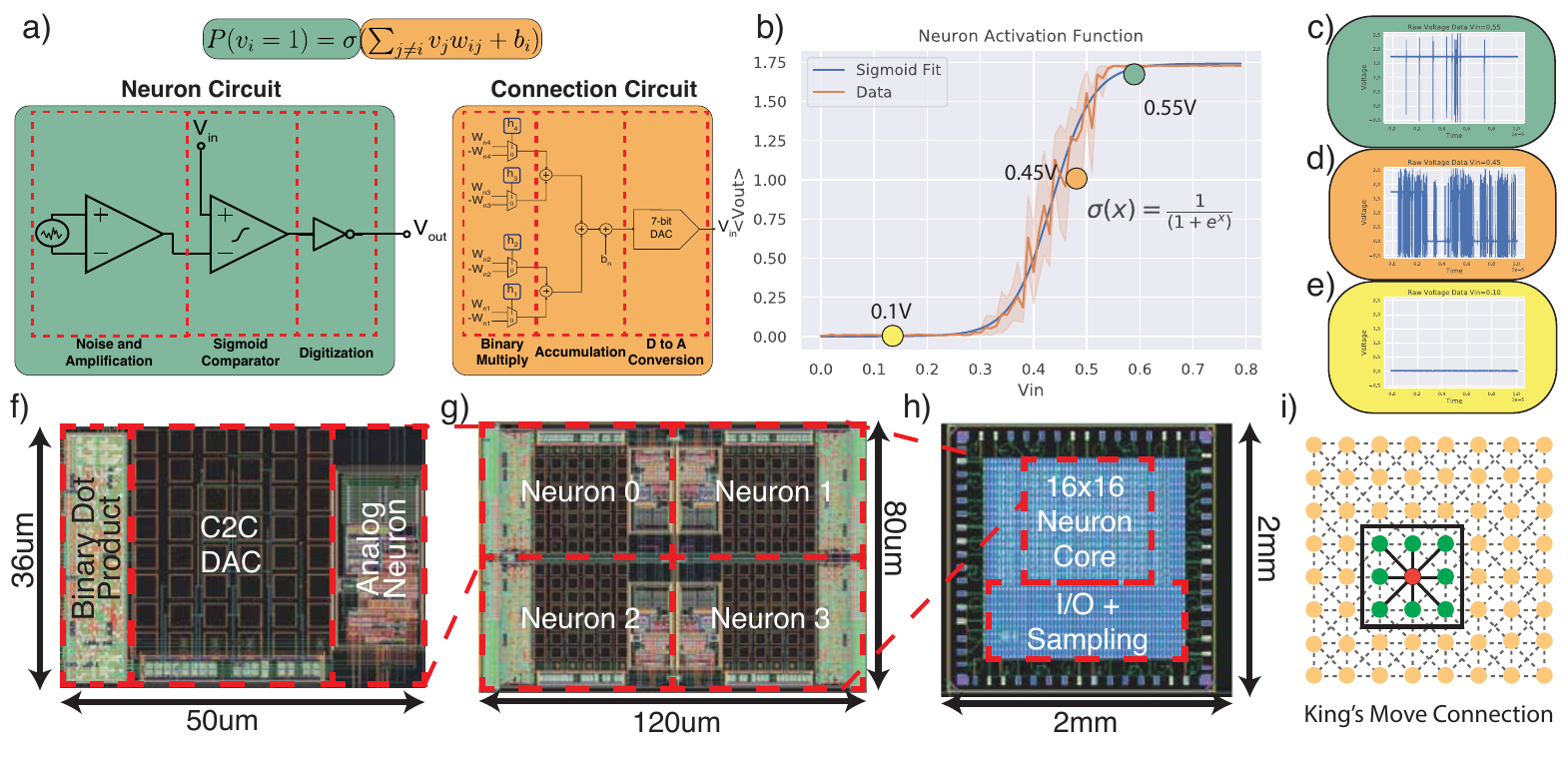}
\par\end{centering}
\caption{\label{fig:hardware} \textbf{ Description of Hardware Design for the PASS system} \\ 
{\textbf{(A)}} A circuit diagram of the neuron and connection circuitry. The neuron circuitry is formed 3 parts, a noise and amplification using amplified shot noise from a diode, a sigmoidal comparator to produce the activation, and a digitization to binarize the output. The connection circuitry is a digitla binary dot product engine, and a digital to analog conversion. 
{\textbf{(B)}} The individual neuron circuit produces an activation function that directly matches the expected sigmoidal activation function. We see that the activation functions well over a wide range of input voltages. This activation was directly extracted from the silicon outputs.   Error bars on the sigmoid data show 95\% confidence interval estimates for the activation function over 100 runs. 
{\textbf{(C)}, \textbf{(D)}, \textbf{(E)}} Raw voltage waveforms for 3 voltages (0.1V, 0.45V and 0.55V) from the sigmoidal activation. These waveforms show how the input voltage modulates the average output value, without the interaction of any input clock signal. 
{\textbf{(F)}, \textbf{(G)}, \textbf{(H)}} Post-layout images of the chip at various levels of integration. The individual neurons are composed of the binary dot product, DAC and analog subsystems, which are then integrated into small clusters, and finally into a 16x16 neuron core. The neuron core consumes 1mm x 1mm of die area, and the peripheral circuitry, I/O and fill consumes the rest of the 2mm x 2mm chip area. 
{\textbf{I)}} The neurons communicate with their neighbors on a kings move graph, where each neuron is connected to its nearest neighbor, as well as diagonal. 
}
\end{figure*}

\clearpage
\begin{figure*}
\begin{centering}
\includegraphics[width=0.88\linewidth]{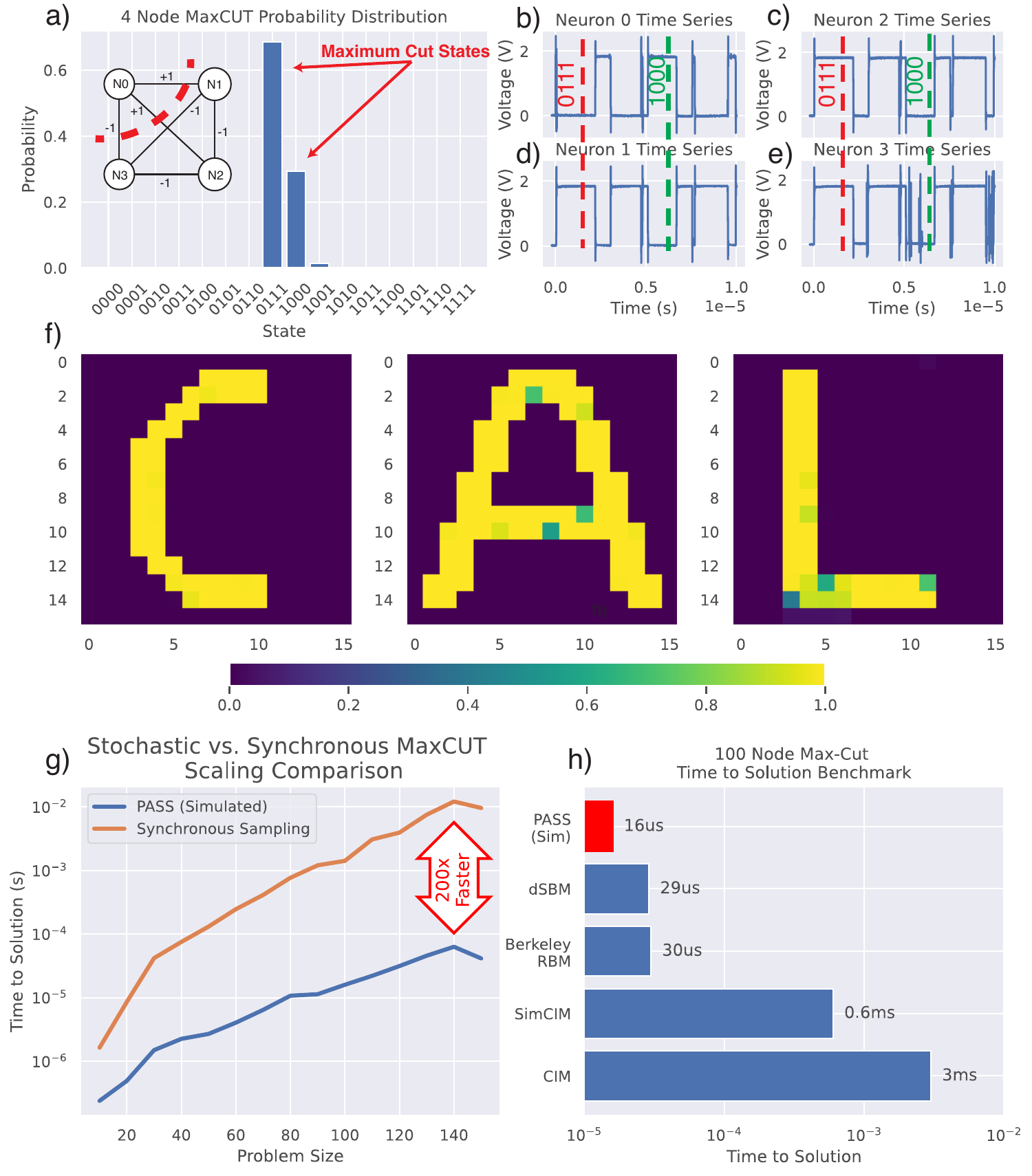}
\par\end{centering}
\caption{\label{fig:maxcut} \textbf{Optimization Tasks} \\
\textbf{(A)\/} Probability Distribution of MaxCUT problem, sampled from Time Series on right. Distribution shows maximums at the two solutions for the MaxCUT problem shown in the figure inset. 
\textbf{(B)\/},\textbf{(C)\/},\textbf{(D)\/},\textbf{(E)\/} Time Series showing fluctuations between correct states in \unit{\us} timescale. Each node spontaneously switches states states and influences the state of neighboring nodes.
\textbf{(F)\/}A MaxCUT problem encompassing the full stochastic core, with the ability model arbitrary problems. The ground state is encoded to spell out the letters C, A, and L, which the system finds with probability approaching 1. 
\textbf{(G)\/} Scaling simulations of the asynchronous PASS system vs. a Synchronous System on the fully connected MaxCUT problem running at the same effective clock frequency. The asynchronous PASS system shows 200x improvement at 150 nodes, with a clear scaling improvement due to usage of the asynchronous system. 
\textbf{(H)\/} Performance Comparison of simulated PASS system compared to state of the art systems on the 100 Node MaxCUT problem (data taken from \cite{Patel2020LogicallyFactorization, Goto2021High-performanceMechanics}). The PASS simulations show the ability to solve problems 2x faster than the fastest alternate solver. 
}
\end{figure*}

\clearpage

\clearpage

\clearpage


\newpage

\bibliographystyle{Science}

\bibliography{references.bib}

\begin{scilastnote}
\item [Acknowledgements] 
The authors would like to thank Vivek Sridhar for helpful conversations on animal decision making. The authors would also like to thank Pratik Brahma and Adi Jung for suggestions and research directions. 

\item [Funding] 
This work was supported by ASCENT, one of six centers in JUMP, a Semiconductor Research Corporation (SRC) program sponsored by DARPA.

\item [Author Contributions] 
S.P. P.C. A.D. and S.L. Designed and implemented the PASS accelerator ; S.P programmed PASS and did initial characterization.;  C.G. performed additional characterization including power analysis; S.P and S.S co-wrote the manuscript; S.S supervised the research. 
All authors contributed to discussions and commented on the manuscript. 

\item [Competing Interests] 
None declared. 

\item [Correspondence] 
Correspondence and requests for materials can be addressed to either S.P. (saavan@berkeley.edu) or S.S. (sayeef@berkeley.edu). 

\item [Code and Data Availability]
Code and Data will be made available on reasonable request by emailing S.P. or S.S. 
\end{scilastnote}

\section*{List of Supplementary Materials}
\begin{itemize}
    \item Materials and Methods
    \item Supplementary Text
    \item Supplementary Figures S1-S7
    \item Supplementary Tables S1-S2
\end{itemize}


\clearpage
\newpage
\section*{Materials and Methods}

\section*{Further Details on Hardware Design}

Each individual neuron amplifies shot noise to create a stochastic clock based on a Poisson process. This update scheme creates an asynchronous neuron that continuously and stochastically computes on the synapse output voltage. As there is no inherent clock for computation, the neuron speed is characterized by the autocorrelation of the Poisson process that governs it; the faster the autocorrelation decays to 0, the faster the neuron computes. Further details of characterization of the autocorrelation and fitting to simulation parameters can be found in the Supplementary materials. Ultimately, this speed is determined by the bandwidth of the amplifiers and size of the noise signal, both of which improve with smaller features, process nodes, and improved design. For this reason, we expect this type of circuit to improve rapidly as technology nodes scale, with ~\si{Ghz} speed being possible with more modern processes. 

The neuron synapse is a multiply-accumulate operation, which is simplified by the use of binary activations with a MUX to perform multiplication. The neuron weights are stored in a distributed memory system addressed by a shift register running through the core. This allows the architecture to overcome problems arising from the von-Neumann bottleneck, as weights are stationary throughout computation. The synapse is digitally synthesized to mux and accumulate signals from adjacent neurons and pass the 7-bit answer into the DAC for conversion to an analog value. The synapse is a fully combinational logic cell, to ensure all neuron computation is done in a clock-free manner. 

Data from a neuron core is sampled at a fixed clock provided by an FPGA host system. We note that although the output of each neuron is asynchronous, and computing is done in an analog-mixed signal manner, the neuron sampling circuitry does not rely on any complex Analog to Digital conversion schemes, and only uses a flip-flop synchronizer circuit. This reduces the overall power consumption and area while simplifying design of the overall system. The FPGA host communicates with the PASS chip to program weights and stream data out of the sampling core.

\subsection*{Noise Source}

The noise source is the heart of the main neuron circuitry. The noise source has a few major requirements for efficient operation. Firstly, there's a need for independence between neurons, ensuring that noise sources are uncorrelated. This design criterion is crucial to minimize interference and enhance the system's overall reliability. Additionally, the output is required to have a peak-to-peak voltage greater than 5mV, which is necessary to provide sufficient signal for the gain stage. Another key requirement is that the noise spectrum should exhibit the characteristics of white noise, resulting in a constant power spectral density to allow for understood probabilistic operation. Lastly, the design must be easy to layout, taking into account the constraints of the manufacturing process. This final point underscores the importance of practicality and feasibility in the design process, ensuring that the circuit can be efficiently manufactured and integrated into larger systems.

The independent neuron constraint is accomplished by having a self biased configuration, where a current mirror load is given to a tie down diode which creates the noise source. To maximize the noise output, the noise diode sees a high impedance load through the drain of the PMOS transistor. The combination of small devices and high impedances (low overall current) means that this system produces sufficient noise to be amplified by the amplification scheme. 

The small size (minimal area diode) allowed for a large noise generation and minimal filtering capacitance. The one issue with this configuration was that the reverse biased diode did not provide exactly a white noise spectrum, which changes the autocorrelation of the final system. This can be attributed to the presence of both a transistor and diode, while the diode may provide a shot noise spectrum, the transistors noise spectrum is a pink noise source. 

\subsection*{Buffer Stage}

The buffer is meant to provide a low input capacitance stage (to prevent filtering of the input noise) while also supplying the necessary low output impedance to drive the gain stage. The buffer was designed with a low output impedance, allowing it to not load the subsequent gain stage which relies on resistive feedback to create signal gain. To accomplish the specification of low output impedance, we chose to use a ``super source follower" configuration shown in Figure \ref{fig:full_circuit}. 

\subsection*{Noise Amplifier}

The noise amplifier is a 2 stage, single-ended amplifier with internal feedback whose goal is to provide a reasonable gain with low enough variation, such that the array of 256 neurons has switching behavior in all devices. The speed and gain of the noise amplifier set the overall speed of the circuitry and are critical to the neuron performance.  Overall the small size of the transistors meant that secondary techniques were necessary to get the amplifier process variation to 3 sigma performance levels and to have a high enough yield to be usable in an array format. These techniques are described below in the discussion on process variation. 

\subsection*{Sigmoid Generation and Output digitization}

\begin{figure*}[!ht]
\begin{centering}
\includegraphics[width=\linewidth]{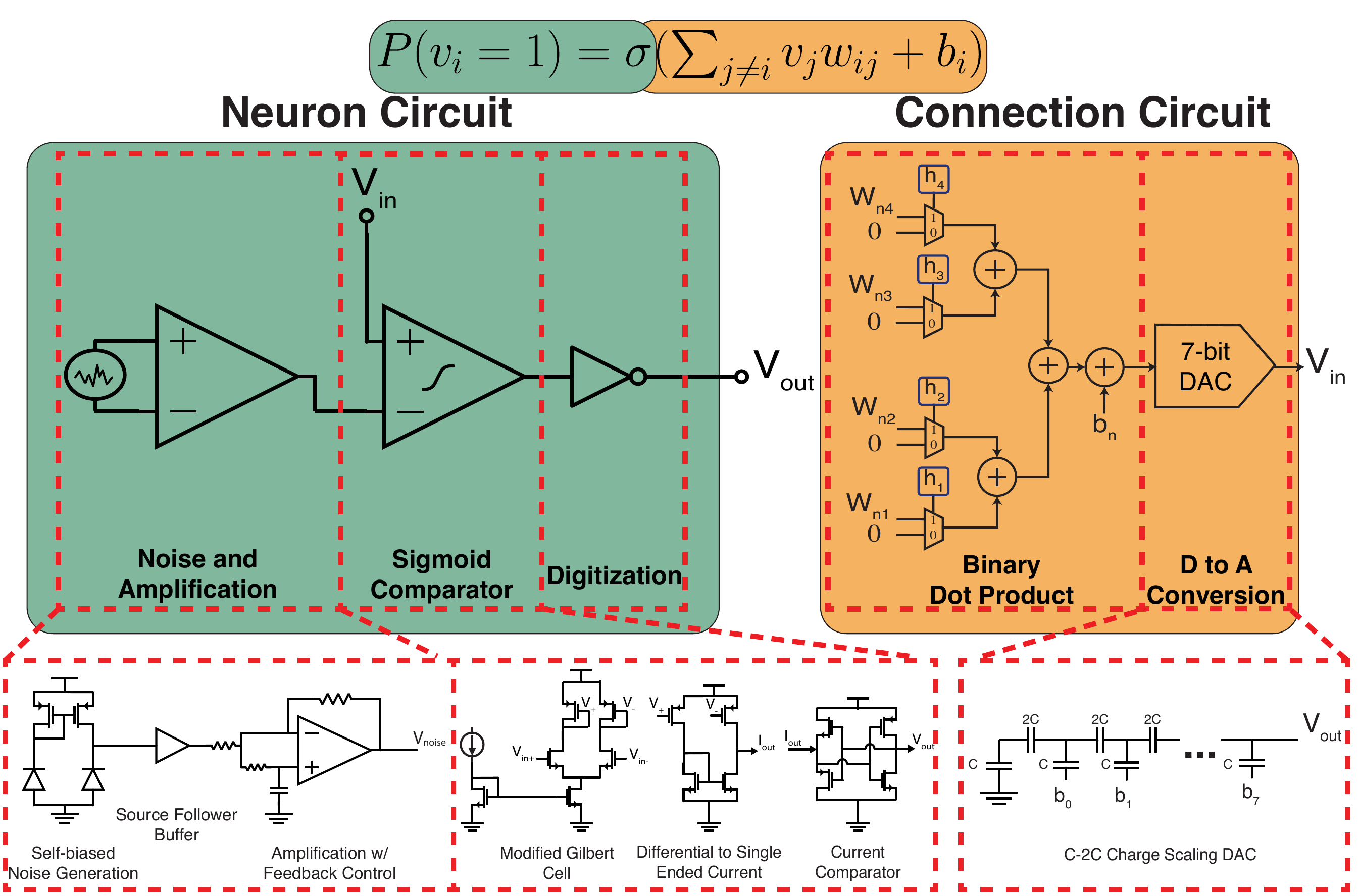}
\par\end{centering}
\caption{\label{fig:full_circuit}\textbf{Circuit diagram of each component of the neuron circuit. }  \\
CMOS implementation of a PASS neuron. The neuron produces a random bit stream biased by the input voltage. Connections between neurons are implemented digitally \\
\textbf{Neuron Circuit:} The Neuron Circuit is divided into three parts, the noise and amplification, the sigmoid comparator and the output binarization/digitization. The Noise is produced by a reverse biased diode, which is put through a buffer and amplification circuit. The noise sources are self-biased to ensure no correlation between neurons. The buffer is a super source follower system which ensures the amplifier does not load the noise generation circuit. The amplifier has feedback control and bias control which allows us to tune its speed and gain after fabrication. The Sigmoid Comparator uses a modified CMOS gilbert cell, which then goes through a current comparator circuit to output an output voltage proportional to the sigmoidal difference between the noise and the input voltage. This then goes into a digitization buffer which drives the asynchronous digital output for adjacent neurons. \\
\textbf{Connection Circuit:} The connection circuit is a two part circuit composed of a digital binary dot product and a 7-bit digital to analog conversion. The binary dot product is a purely combinatorial block, which allows the system to use digital weights stored in registers to do the communication. The digital to analog conversion is done using a C-2C architectures which yields a high speed, low power system. }
\end{figure*}

After the amplified noise signal is generated by the OTA stage, the signal is sent into the sigmoid comparator. The sigmoid comparator is responsible for comparing the input noise signal to the signal coming from the synapses and DAC. It can be understood as doing the probabilistic activation function which can be seen as $\sigma(x) > R$ where $R$ is the random signal coming from the synapse. The sigmoid circuit is divided into 3 parts shown in Figure \ref{fig:full_circuit}. 

The first part is a modified gilbert cell which formulates the activation function shape \cite{Gray2009AnalysisCircuits}. We note that a gilbert cells usually would either use BJT input transistors or have the input MOSFET transistors biased in subthreshold. In our case, our process flow did not offer BJT transistors with enough current per area to be area efficient, and biasing in sub-threshold meant that we would sacrifice too much speed for the sigmoid. We sacrificed a perfect sigmoid shape to have the speed needed for operation. With proper tuning, we found that the activation shape was sufficiently close to sigmoidal for our applications. 

The second and third part of the circuit take the differential current output from the gilbert cell and convert it to a single binary output voltage. We use a current comparator composed of a bistable inverter first presented in \cite{Freitas1983CMOSCircuit}, which gives a sufficiently fast output circuit. The final voltage output is passed through an inverter to ensure proper drive strength and to force the output to be purely binary. 

\subsection*{Synapse Design}

The neuron synapse is a multiply-accumulate operation, which is simplified by the use of binary activations with a MUX to perform multiplication. The neuron weights are stored in a distributed memory system addressed by a shift register running through the core. This allows the architecture to overcome problems arising from the von-Neumann bottleneck, as weights are stationary throughout computation. The synapse is digitally synthesized to mux and accumulate signals from adjacent neurons and pass the 7-bit answer into the DAC for conversion to an analog value. The synapse is a fully combinational logic cell to ensure all computation is done in a clock-free manner. 

\subsection*{Digital to Analog Converter (DAC)}

The Digital to Analog Converter (DAC) takes the digital accumulation from the synapse cell and converts it to an analog voltage which can then be used by the analog neuron cell and the sigmoid comparator. The DAC uses a C-2C topology which allows for area efficient conversion of digital to analog output values. The only problem with these C-2C topologies is charge leakage off of the output node as time goes on. We find this means there is a time window for computation to occur before it is invalid. 

The C-2C topology, and other capacitor based DAC topologies, have the added benefit of being very power efficient, with no leakage current, and power only used to charge the capacitors during switching. Along with this, the speed of the DAC is governed by the inverters driving the bit lines of the DAC, with switching speeds under $<10$\si{ps}. 

\subsection*{Sampler and I/O Design}

The network is laid out in a regular grid with nearest-neighbor connections. To derive network statistics, we sample each neuron’s binary output at a fixed clock frequency and stream out data off chip. We use a flip-flop synchronizer to ensure low probability of metastable states as we move from the asynchronous domain to the sampling synchronous clock domain. 

The sampler and I/O is designed to be as simple and robust as possible to focus on design of the analog neuron core and to ensure the greatest chance of success of the system. The design of the system is composed of 3 components, as described below:

\begin{itemize}
    \item Neuron State Sampler: This circuit samples the neuron's binary output at a rate of $>100$\si{Mhz} and shifts the output down the sample column to hold for eventual readout. 
    \item Chip Configuration: The configuration chain is responsible for setting the weights, biases, digital trims, and sampler settings for the full chip. It is implemented as a long shift chain. 
    \item Data Readout: The readout circuitry takes the neuron states that are held in an SRAM buffer and sends them out at an I/O friendly speed of $\approx 10$\si{Mhz}. 
\end{itemize}

\subsubsection*{Neuron State Sampler}

As the neuron states are binary output, no complex ADC circuitry is necessary to sample the output state of the system. Instead, a 3 register flip-flip synchronizer is used to ensure the circuitry does not enter a metastable state and to move from the asynchronous domain to the sampled clock domain. A $300$\si{Mhz} clock is sent to each neuron, with the neuron state being sampled once every $k$ cycles, where $k$ is the number of rows we wish to sample at a given time. This means that each neuron would be sampled at a rate of $\frac{300}{k}$ \si{Mhz}, allowing us to sample a few rows at a very fast rate to get more accurate statistics, or to sample the full chip to get less granular statistics but over a larger chip area. This design decision helped relax timing requirements for movement of the sampled chip data off the chip.  This $k$ parameter is set in the configuration chain at set up time. Once the sample is taken, it is shifted along the sample column and finally held in the SRAM buffer. 

The SRAM buffer for the Neuron State sampler is a 136kBit SRAM cell with 17 bit lines, which is able to hold 8192 row samples at a given time. The extra bit per row is meant to hold a ``fingerprint" bit, which allows us to synchronize the origin of each sample after samples are streamed to the computer for analysis. If we are sampling the full internal cluster, this means we are getting 512 samples from each individual neuron, or $\approx 440$ \si{us} of data sampling the full cluster with a $300$ \si{Mhz} clock. 
 
\subsubsection*{Configuration Chain}

The configuration chain holds all of the weights and biases, as well as sampling and trim information for the neuron core. Each neuron has 74 configuration bits, with 8 bit weights for each of the 8 neighboring neurons, 8 bits for the bias, and 2 bits to clamp the output to either 1 or to 0. Along with this, the configuration chain holds the 7 bit configurations for 16 current trimming circuits for the amplifiers and 16 current trimming circuits for sigmoidal activation circuitry. The last piece of the configuration chain is a 3 bit sampling configuration register, which corresponds to how many rows of the chip core we wish to sample at a given time (the settings are 1 row, 2 rows, 4 rows, 8 rows, or 16 rows to sample at a given time). 

The configuration chain is updated on a much slower 1 \si{Mhz} clock. The configuration chain is intentionally designed with a slower clock as the configuration data must snake across the entirety of the chip. This also allows for sufficient timing slack to optimize for other pieces of the system design. 

\subsubsection*{Data Readout}

Data from the SRAM buffer is read out at a slower I/O clock rate (20 \si{Mhz}) and shifted out in a Parallel-In-Serial-Out (PISO) fashion through a single General Purpose I/O (GPIO) pin. This design decision was done to minimize use of I/O pins when necessary, as it was deemed that I/O speed was not necessary for this proof of concept system. Additionally fewer I/O pins mean that more chip outputs can be dedicated to power, ground, and test outputs.

\subsection*{System integration}

The system is finally integrated in a mixed-signal fashion to produce the final $16$ x $16$ array of neurons. An image of the post-layout chip is shown in Figure \ref{fig:chip_photo} and \ref{fig:hardware} where the integrated system is shown at a neuron level, a small scale integration level, and at the full chip level. We further are interested in the power, performance, and area of the full chip system.

The 256 neuron core takes up a $1\si{mm}$ x $ 1 \si{mm}$ in the center of the core, with the majority of the rest of the chip being fill area and a small area devoted to I/O and the sampling circuitry. When analyzing the area of the chip, we find that the DAC has the highest area contribution within the cluster, while the binary dot product engine and analog neuron are smaller components. This is due to the necessity for large capacitors to combat layout parasitics. A full area accounting of a single neuron system is given in Table \ref{tab:single_neuron_area}. 

When analyzing the power of the full system, the analog neuron takes the majority of the power, specifically the Noise Amplifier and Noise Buffer, both of which take $\approx 30 \si{uA}$ of current when the full neuron taking a total of $\approx 80 \si{uA}$ of current. We find that the digital accumulation of the system takes a much lower average current amount, as it does not run continuously, unlike the analog sub-systems. The full accounting of power consumption of a single neuron system can be found in \ref{tab:single_neuron_b0_power}. 

As discussed previously, the speed of switching in the system is largely governed by the performance of the analog Noise Amplifier and noise generation system. To achieve faster performance, a higher bandwidth and higher gain amplifier would have to be designed, which in turn would cause a larger power draw. If we are willing to compromise on power of the system, we can decrease the bandwidth of the noise amplifier (by operating it in sub-threshold regime, for example) for next generations of design.

\subsection*{Integration with Classical Computer}

To integrate with a classical computer, the PASS system was mounted on a chip-on-board configuration style. The PASS chip has all digital test signals outputted through the carrier board, while all analog signals (including test neuron outputs) are able to be directly probed via an analog SSMB connector. The carrier board interfaces with an Agilent DSO2024 oscilloscope for analog signal where the signals are then output and interpreted to a traditional computer. 

\begin{figure*}
\begin{centering}
\includegraphics[width=\linewidth]{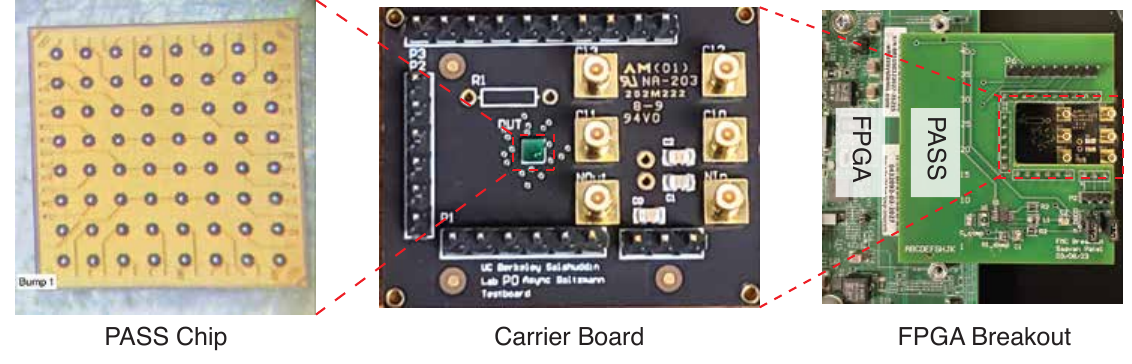}
\par\end{centering}
\caption{\label{fig:chip_photo}\textbf{Micrograph of PASS chip, carrier board and integration with FPGA system}  \\
\textbf{Left:} Photograph of the ball grid array and redistribution layer routing for the PASS chip. The system uses a chip on board architecture for connection to a carrier board. \textbf{Center:} Carrier board for breakout connections of the PASS chip. Analog signals are put through SSMB connectors on the right, while digital signals are put through standard connector pins. \textbf{Right:} Carrier board with FPGA breakout, communicating with Xilinx Ultrascale+ VCU118 evaluation board through FMC connectors. }
\end{figure*}

For more complex integration, the carrier board is combined with an FPGA breakout board which interfaces with a Xilinx VCU118 FPGA evaluation board. This FPGA board supplies all necessary signals (clocks, data handshakes, and computer integration) to run the PASS system at full speed. This then takes the generated data and transmits it back to the computer for study and integration. The full system integration and carrier board is show in Extended Data Figure \ref{fig:chip_photo}.

\subsection*{Controlling Process Variation}

As with most analog designs, process variation is a major challenge when implementing an array of a large number of neurons. Process Variation is more pronounced with analog sub-systems as we expect the output to be a precise value rather than a binary output. To combat design variation, we employed 5 major methods at different levels of the design hierarchy: Over-sized amplifiers for reduced transistor variation, Auto-Zeroing for offset cancellation, digital trimming of amplifier and neuron current, post-fabrication offset correction during programming, and modification of the supply voltage. 

As is generally understood in analog systems, larger transistors have less pronounced transistor to transistor variation (and less variation across multiple dies) \cite{Gray2009AnalysisCircuits}. Defects present in larger transistors are less likely to destroy the entirety of the transistor performance and can be better managed. In the analog components whose variation needs to be well controlled, and is not controlled by other methods (such as the Noise Buffer and sigmoid), we have made efforts to make the system as large as possible to reduce neuron to neuron variation. 

\begin{figure*}
\begin{centering}
\includegraphics[width=0.88\linewidth]{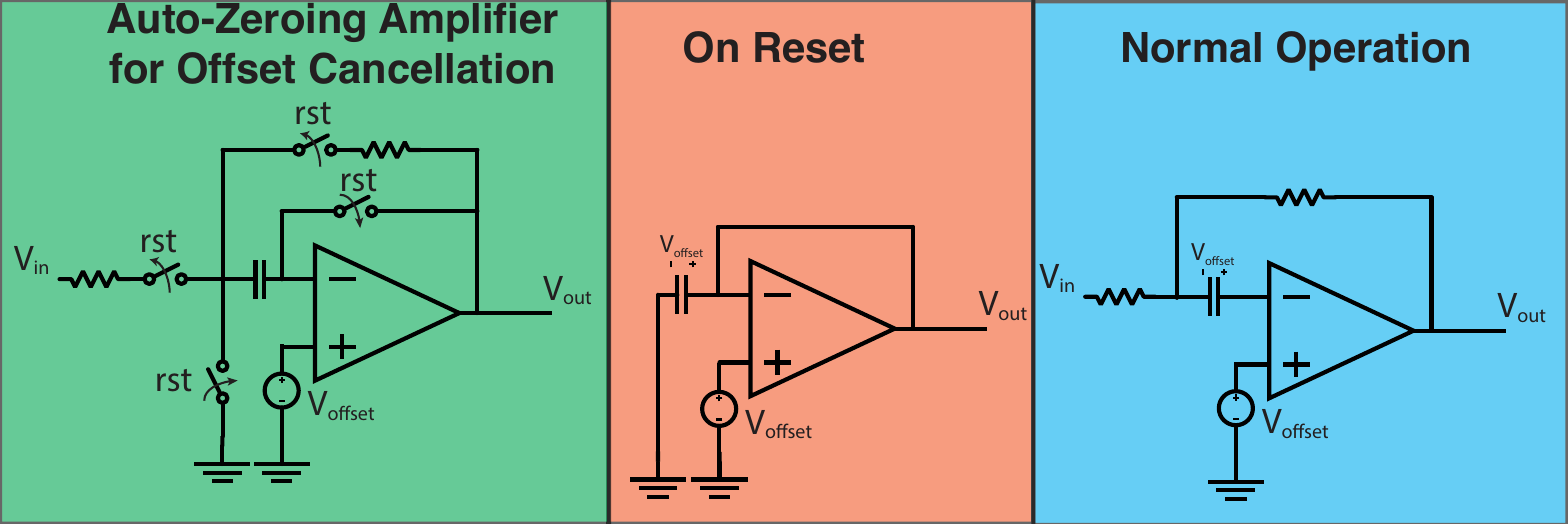}
\par\end{centering}
\caption{\label{fig:autozero}\textbf{Autozeroing circuit for correction of sigmoid output}  \\
\textbf{ Autozeroing circuitry for the Noise Amplifier Circuit. Left: the full neuron circuitry, including switches which are flipped open or closed on reset. Middle: When the amplifier is reset, it is put into unity-gain to cancel out the effect of the input offset voltage. Right: On normal operation, the reset circuitry does not effect regular operation, and it is able to operate in regular negative feedback configuration. }}
\end{figure*}

The performance parameters of the Noise Amplifier have the greatest effect on the overall neuron performance. The speed of the noise amplifier sets the speed of the overall neuron (along with the spectrum of the noise source), the amplitude of the noise amplifier sets the width and shape of the activation function, and the common mode output voltage of the amplifier sets the center of the sigmoidal activation. For this we have gone through great lengths to decrease the variation of this system. To offset any input offset voltage, we have implemented an Auto-Zeroing circuit \cite{Gray2009AnalysisCircuits, Enz1996CircuitStabilization}. Before computation is conducted, a reset signal is broadcast to every neuron, which cancels any drift in the amplifier. The possible length of computation is partially set by the leakage off of the internal capacitor node of the auto-zeroing circuit, which causes some drift in the common mode output of the amplifier over the course of $\approx 1$\si{ms}. The topology of this circuit is shown in Extended Data Figure \ref{fig:autozero}.

To further reduce the effect of the Noise Amplifier variation, we have introduced a digital trimming circuit for both the noise amplifier and the sigmoidal activation function. The digital trimming circuit is part of the configuration chain and modifies the input current to groups of 16 neurons together. The effect of the digital trim of current on the noise amplifier is to change the amplifier gain (which is itself controlled by the feedback resistors of the amplifier), and to control the speed of the amplifier. If variation of the amplifier causes its gain to become too low or the neuron to switch too slowly or too quickly, we are able to trim it using this method.

\begin{figure*}
\begin{centering}
\includegraphics[width=0.88\linewidth]{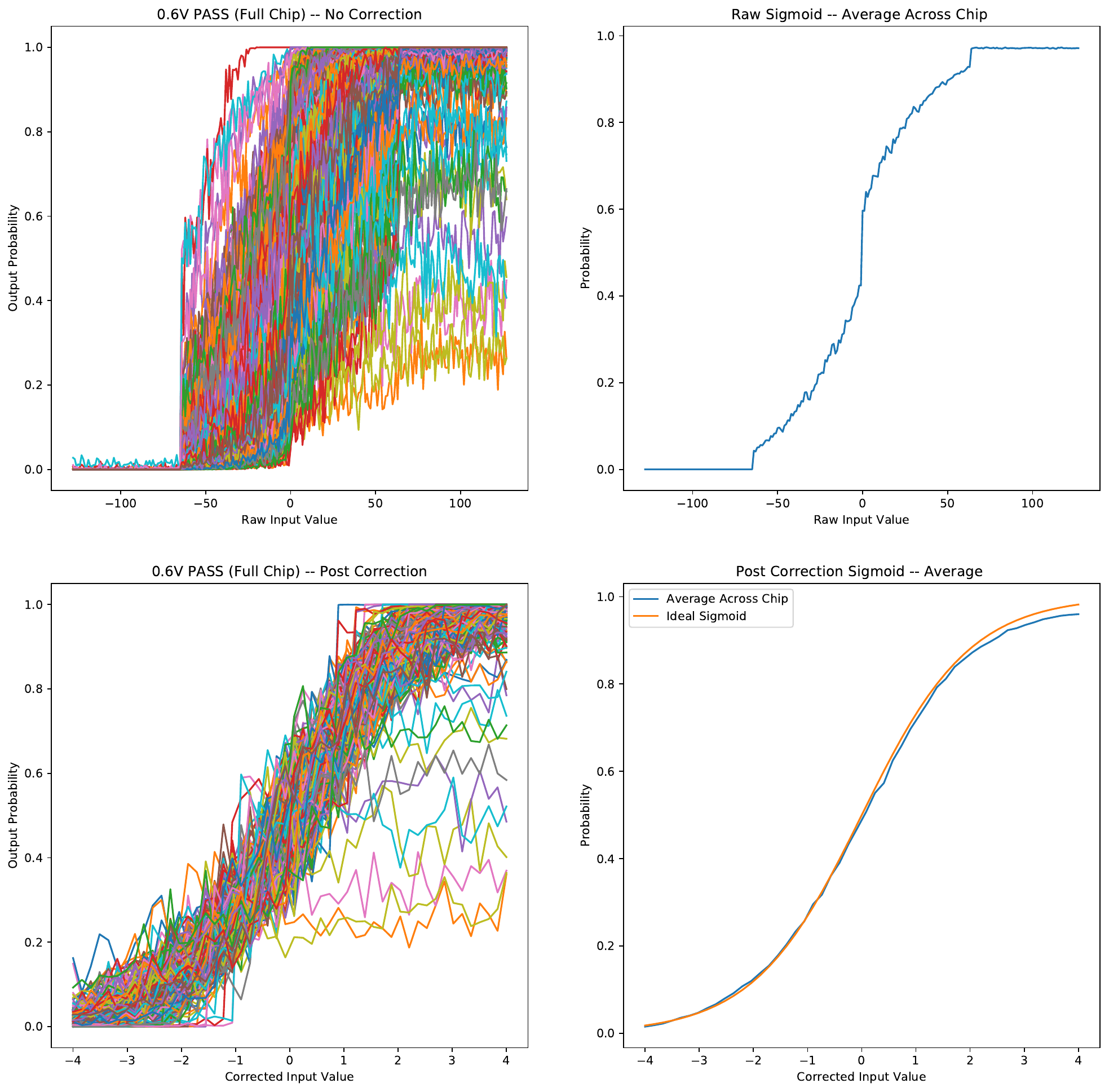}
\par\end{centering}
\caption{\label{fig:sigmoid_correction}\textbf{Correction of the Sigmoidal activation post-fabrication by fitting sigmoids to each neuron individually}. \\
Top Left: A plot of all neuron activations across a chip plotted on top of eachother. There is a lot of variation in activation between neurons. Top Right: The extracted average activation function across the chip. This activation is close to sigmoidal, but still requires fitting. Note that althought it appears that the average saturates to 1 across the chip, it is slightly below 1 due to faulty neurons. Bottom Left: After doing a linear fit of each neuron activation, we see that the neurons have much less variation across the chip, and generally have a closer to ideal activation. Bottom Right: When averaged across the chip, the average activation function becomes almost exactly the ideal sigmoidal activation, which is plotted on top of it for additional information. }  
\end{figure*}

To reduce variation after fabrication, we are able to extract activation data from each individual neuron and do a correction step before programming the neuron array with the problem of interest. This is done by a linear correction term described in Equation \ref{eq:sigmoid_correction} below. The sigmoid is individually characterized by incrementing the bias input code from -127 to +127 (the smallest and largest codes for the neuron). After this, a sigmoid with two linear correction parameters is applied to each individual neuron, which allows each neuron to have a different linear offset and slope of activation. The effect of implementing this correction is demonstrated in Figure \ref{fig:sigmoid_correction}, where we see that the average sigmoid variation is drastically decreased. Note that even with the correction, we see some neurons still behaving incorrectly (seen as the neurons which do not saturate to a probability of +1 on largest input). These neurons are ``dead" and we avoid programming them. 

\begin{equation}
    \label{eq:sigmoid_correction}
    \sigma(x) = \frac{1}{1 + e^{-(a(x-b))}}
\end{equation}

\begin{figure*}
\begin{centering}
\includegraphics[width=0.88\linewidth]{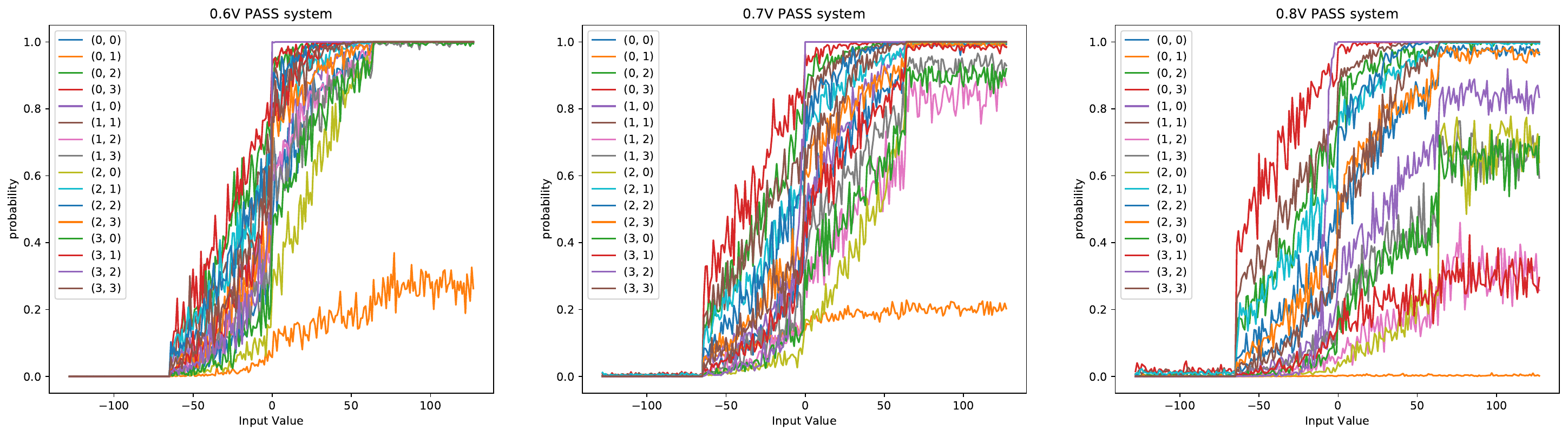}
\par\end{centering}
\caption{\label{fig:voltage_variation}\textbf{Dependence of variation on input voltage }  \\
Analyzing neuron behavior as a function of supply voltage. Looking at the same neurons going from 0.6V to 0.7V to 0.8V, we can see that as voltage increases, the variation amongst neurons increases as well. We can attribute this to the amplifier circuitry overpowering the sigmoid generation circuitry causing the neuron to never fully saturate to the +1 state. }
\end{figure*}

The last method of reducing variation in the chip is to modify the supply voltage of the analog core. This is somewhat of a non-intuitive result but proves to be an important way of forcing the neurons to behave in a predictable way. When we decrease the supply voltage of the neuron, the noise amplifier is effected most by the change in behavior. This means that the gain and speed parameters of the amplifier start to degrade. When the gain degrades, the output sigmoid will become ``narrower" due to it becoming easier for the output of the DAC to be larger than the output of the noise amplifier. In the case of this PASS system, the noise amplifier produces more noise than originally specified at nominal supply voltage of 0.8V,causing many of the neurons to never fully saturate to the +1 state. Thus, although the system is slower at 0.6V, all experiments for complex problems are conducted at this reduced supply voltage to maintain accuracy and predictability. These results are visually shown in Extended Data Figure \ref{fig:voltage_variation}.

\subsection*{Delay Analysis}

Delay between neurons can have a large effect on performance of the overall system. To understand the affects of delays between neurons and how to scale the delay as compared to the autocorrelation and speed of other circuit components, we have done a series of experiments to understand this. 

There are two components to delay, a probabilistic delay which comes from the nature of the neuron to switch at a random time based on its inputs, and a deterministic delay which has to do with standard circuit delays such as routing delays and delays through circuit components. There are two ways that we characterize the delay of each of these components. 

To understand the probabilistic delay, we characterize the system using the autocorrelation function of its output. The probabilistic delay can be understood as the expected amount of time that a state will remain in the current state. Mathematically, we represent the probabilistic delay using the autocorrelation function shown below. 

\begin{align}
    \label{eq:acf}
    ACF(\Delta t)  & = \frac{E[(x(t + \Delta t) - E[x(t + \Delta t)])(x(t) - E[x(t)]]}{\sigma_{x(t + \Delta t)}\sigma_{x(t)}} \\
    & = \frac{E[(x(t + \Delta t) - \mu_{x(t)})(x(t) - \mu_{x(t)})]}{\sigma_{x(t)}^2}
\end{align}

When normalized by the variance in this way the autocorrelation function is always $-1 < ACF(\Delta t) < +1$. As the autocorrelation decays to 0, the state at the time step $\Delta t$ ahead in time will be more likely to be uncorrelated with the current time step. As understood in the simulation model, the autocorrelation demonstrates an exponential decay, with the exponential fit parameter representing the average flip rate of the system. The probabilistic delay is the fundamental limiter to faster computation, as it sets how quickly sampling is performed and the chip moves through the state space. 

\begin{figure*}[ht!]
\begin{centering}
\includegraphics[width=0.95\linewidth]{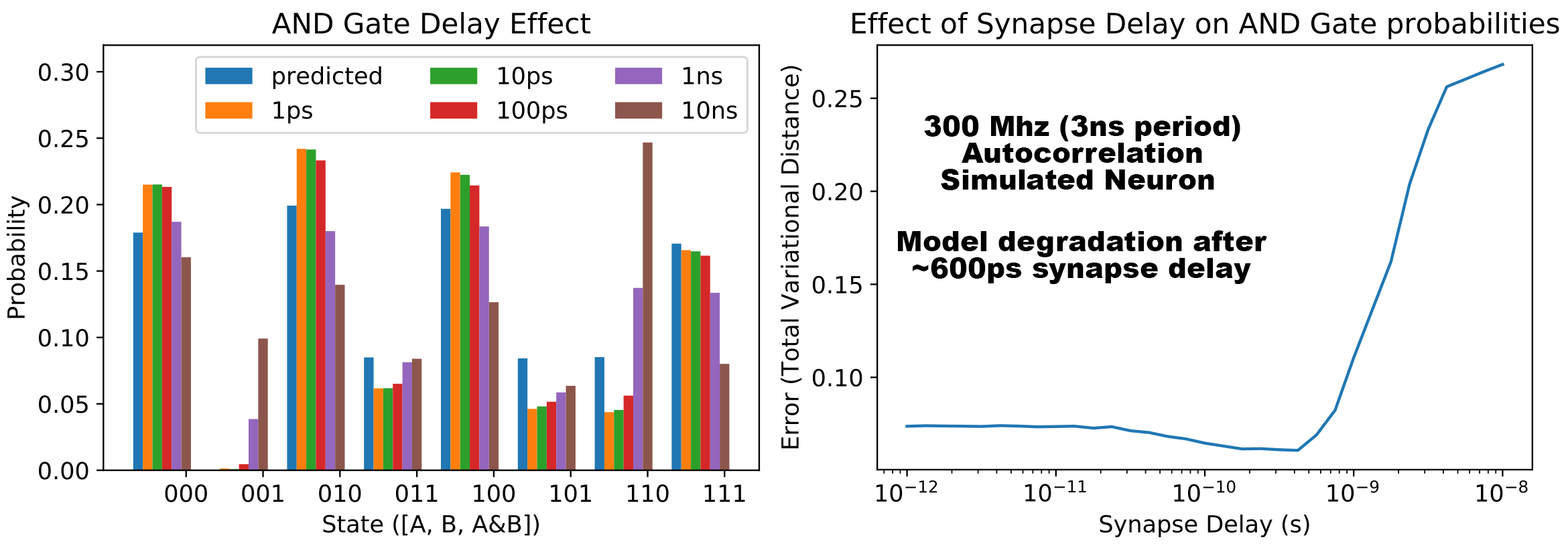}
\par\end{centering}
\caption{\label{fig:delay_sim}\textbf{Analysis of effect of delay on probabilistic performance} \\
Experiments were done on the SPICE neuron circuit, while adding in delays to understand effect of routing and combinational delay. The neuron simulated had a ~300\si{Mhz} autocorrelation which is approximately twice that of the experimented on neuron \textbf{Left:} Probability distributions of an AND gate, with predicted probability distribution vs. distributions measured with various amounts of additional delay. \textbf{Right:} The total variational distance of the predicted distribution vs. the measured distribution on various levels of delay. Significant performance differences seen after additional 600\si{ps} for a 300\si{Mhz} intrinsic clock frequency neuron. 
}
\end{figure*}

To understand the deterministic based delay we have analyzed both simulation and physical data. It is clear that the deterministic delay puts an upper bound on the autocorrelation frequency possible for the system. This can be intuitively understood as neurons need to communicate their information between each other before each of them can compute on the input data, the inputs to the neurons can't be ``stale". To characterize this we performed simulation experiments on the SPICE model of the hardware neuron we are working with as shown in Figure \ref{fig:delay_sim}. In this experiment, we removed all combinatorial delay from the DAC and combinatorial logic and added a variable delay block. We slowly increased the delay and characterized the probability distribution on a simple AND gate model. The predicted probability distribution is one extracted from analytical behavior of the hardware neuron and serves as a reference distribution to compare against. We have seen that the introduction of a ~600\si{ps} delay is the point where the delay starts to significantly alter the probability distribution being sampled from. 

This analysis gives us the below approximate relation between the probabilistic autocorrelation delay ($\tau_{acf}$ and the deterministic circuit delay ($\tau_{circ}$. In general the circuit delay should be 5x smaller than the autocorrelation to not skew the distribution too significantly and cause problems. 

\begin{align}
    \label{eq:delay_const}
    \frac{\tau_{acf}} {\tau_{circ}} > 5 \ \ \ (simulated)
\end{align}

To understand how the actual hardware performs compared to this estimate, we analyzed voltage switching behavior of the free running neuron with a constant input to characterize the autocorrelation. In the case of the PASS system shown here, the $\tau_{acf}$ decay constant shows an average flip rate of $\approx 150 \si{Mhz}$ at maximum speed or a average time between flips of $\tau_{acf} \approx 6.7 \si{ns}$ as seen in Figure \ref{fig:acf}. 

\begin{figure*}[ht!]
\begin{centering}
\includegraphics[width=0.95\linewidth]{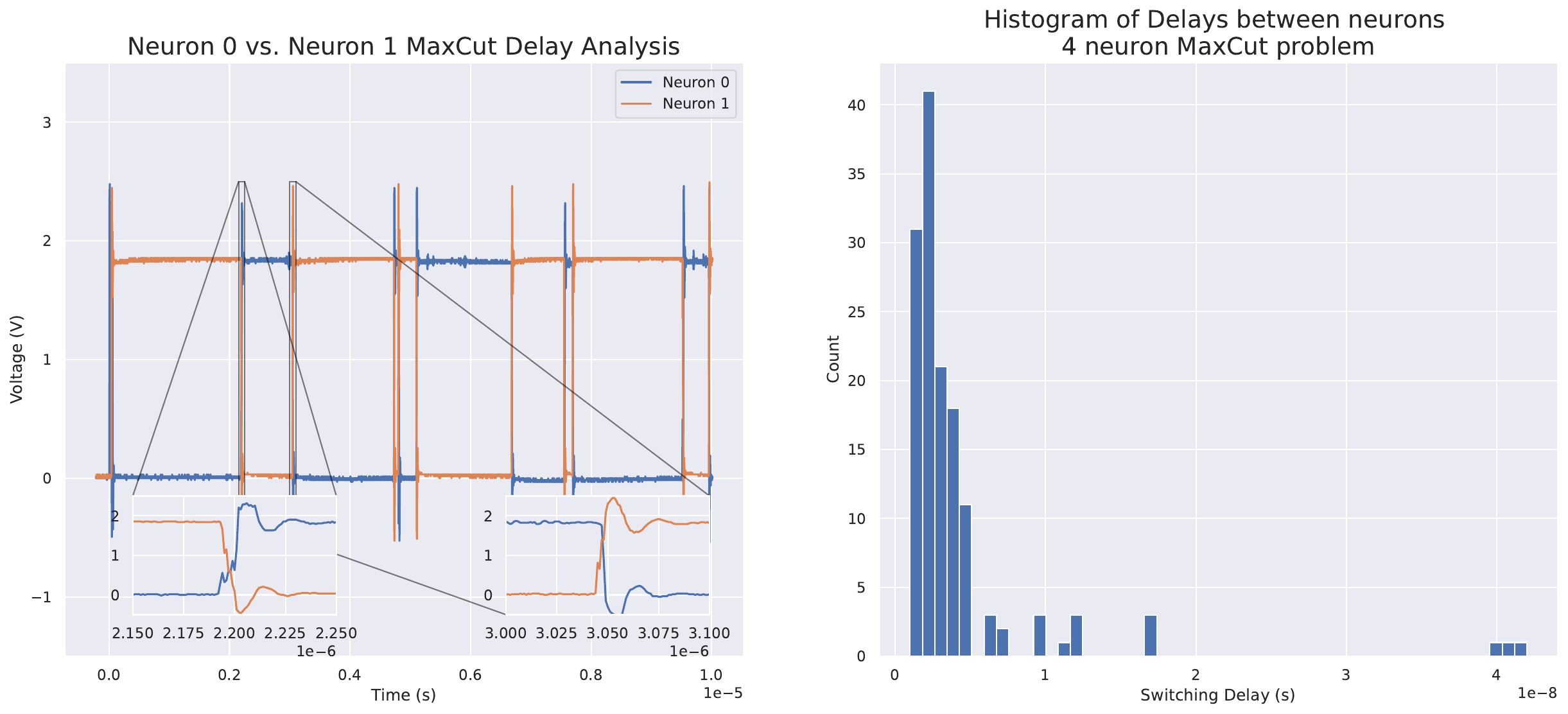}
\par\end{centering}
\caption{\label{fig:delay_hw}\textbf{Analysis of delay in hardware MaxCut problem}  \\
\textbf{ Left:} Time domain switching behavior between two neurons in MaxCut problem, with insets of voltage levels of respective neurons. 
\textbf{ Right:} Histogram of switching delay measured amongst 4 neuron MaxCut performance shown in figure \ref{fig:maxcut} A). Switching delay is measured as the time one neuron takes to drop across 0.9\si{V}} level, and for the connected neuron to rise above the 0.9\si{V} level. Median switching delay is measured as $\approx 2$\si{ns}.
\end{figure*}

To understand the delay of the actual hardware neurons to its neighbors, we used simple probability distributions where we expect the majority of delay to be governed by combinatorial delay. This happens when the weights of the distribution are large, and cause the whole system to flip between two states, as seen in the MaxCut experiment shown in Figure \ref{fig:maxcut} A) where the system flips between the two possible ground states of the solution. When analyzing the voltage signals of this system shown in Figure \ref{fig:delay_hw}, we find that the median delay is $\tau_{circ} \approx 2 \si{ns}$. This yields to the below circuit versus delay ratio.

\begin{align}
    \label{eq:delay_const_real}
    \frac{\tau_{acf}} {\tau_{circ}}  = \frac{6.7 \si{ns}} {2 \si{ns}} \approx 3.3 \ \ \ (actual)
\end{align}

We note that while this breaks the above rule of thumb outlined through simulation, we were still able to get high quality performance and results as demonstrated throughout the paper.  We note that on a further redesign of the system it would be prudent to increase the delay ratio to allow for closer agreement between the input probability distribution and the actually sampled probability distribution. We also note that the PASS chip was designed with the ability to modify the autocorrelation of the neurons on the fly by digitally trimming the current into each neuron. This addition allows the PASS chip to adaptively change this delay ratio as needed in case the output distribution no longer matches the intended values. 

\section*{Simulation Model}

A series of simulation studies were conducted as part of this work to understand the behavior the PASS asynchronous system vs. an equivalent synchronous system. First we'll outline the asynchronous simulation model, then the synchronous simulation model and explain how these can be compared. 

To understand how the asynchronous PASS architecture compares to other state of the art architectures for optimization on these Ising Model architectures, we use a simulation model to project how the system would work on larger problem sets. We first extract activations and fit exponential decay curves to the autocorrelation curves from the hardware and then fit them to the Continuous Time Poisson Process model \cite{Hayes2007AGraphs}. This models the neurons as each having an individual poisson clock, which updates at random times that are modulated by the activation probabilities. This is a similar system model to the Gaussian Machines introduced in \cite{Akiyama1989CombinatorialMachines}. 

The simulation model flow functions as follows. Each neuron is updated according to a poisson random variable that is modulated by its individual neuron update probability. This makes the neuron updates follow an exponential random variable, with distribution shown below in Equation \ref{eq:poisson}. 

\begin{equation}
    \label{eq:poisson}
    P(T > t|s) = e^{-\lambda t}
\end{equation}

This matches physical experiments, where we expect that the autocorrelation for our neurons to follow a poisson process as the noise generation for the underlying neurons are shot noise processes, which are inherently poisson \cite{Gray2009AnalysisCircuits}. 

\begin{equation}
    \lambda = \begin{cases} 
      \sigma(Wv + b)\lambda_0 & s = 1 \\
      (1 - \sigma(Wv + b))\lambda_0 & s = 0 
   \end{cases}
\end{equation}

\begin{figure*}
\begin{centering}
\includegraphics[width=0.88\linewidth]{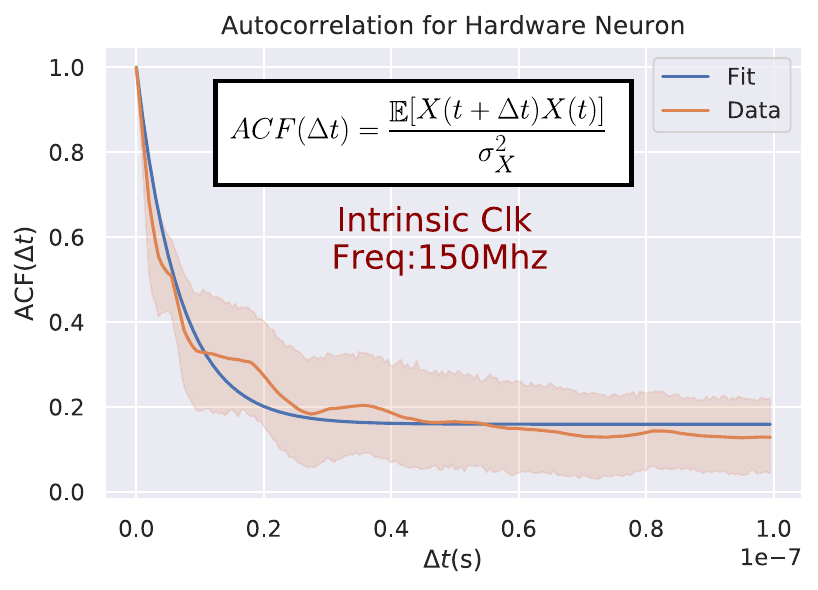}
\par\end{centering}
\caption{\label{fig:acf}\textbf{Autocorrelation fit of PASS chip with switching behavior at maximum speed setting}  \\
Data taken from many runs of the PASS chip at the maximum speed setting. The autocorrelation is fit to a standard exponential decay function, which sets the rate parameter of the poisson process. The rate parameter is extracted to be 150 \si{Mhz} which can be seen from this graph. Additionally we can see a good fit between data and the fit for our system, suggesting that this model is able to reasonably capture the system dynamics. We use this exponential decay fit to calibrate simulation models used to do simulation testing. In this case the $\lambda_0$ value would be $\frac{1}{150Mhz} \approx 6.7ns$ representing the average flip rate of a free running neuron.  }
\end{figure*}

The $\lambda$ value for a neuron is governed by the inherent $\lambda_0$ which is a device parameter extracted from a free running activation function, as well as the input weights and biases to the system. The $\lambda_0$ is characterized by fitting to the exponential decay curve in Figure \ref{fig:acf}. The $\sigma(x)$ is the neuron activation function. For a free running neuron $\lambda_0$ is also the average rate at which the neuron will flip, as it is the poisson process fitting parameter for each update. We can define this parameter as the ``effective clock frequency" for this reason and can use if for comparison to update speeds of continuous systems. This can equivalently be considered as the model for a continuous time markov chain, with two states. 

For the synchronous time system, we use a standard random scan Gibbs sampling framework for simulation of the Ising Model system \cite{Hayes2007AGraphs}. This functions by randomly selecting an element in the Ising Model system, and updating its state based on the states of its neighboring spins. In this simulation, the updates are done with the same rate parameter $\lambda_0$ such that the average update for a neuron in the asynchronous case is the same as the clock speed for the synchronous Gibbs case. This results in a serial update of each neuron one by one, but due to the structure of the Glauber Dynamics and Gibbs Sampling frameworks this is the standard option for updates in an Ising Spin system \cite{Aadit2022MassivelyMachines, Camsari2017StochasticLogic}. We also note that parallel updates are only possible in the case where the graph can be graph colored and sampled \cite{Gonzalez2011ParallelTrees, Aadit2022MassivelyMachines}, which is only possible in the non-fully connected cases. 

\section*{Neural Decision Making models}

Fly decision making is modeled by an Ising Model format as described below in equation \ref{eq:fly_hamiltonian}. In this model, each spin represents a neuron making an individual decision to move towards a target, while the pairwise couplings $J_{ij}$ represent the interaction strength between the neurons attempt to go in opposing directions. 

\begin{equation}
    \label{eq:fly_hamiltonian}
    H(s) = \frac{-k}{N} \sum_{i \neq j} J_{ij}s_is_j \\
\end{equation}

Here $k$ is the number of options each individual has, $N$ is the number of neurons,  and $J_{ij}$ is the interaction strength between neurons. The coupling parameters are set using a cosine geometry shown below in Equation \ref{eq:fly_coupling}. 

\begin{equation}
    \label{eq:fly_coupling}
    J_{ij} = cos(\pi(\frac{|\theta_{ij}|}{\pi})^\eta)
\end{equation}

In Equation \ref{eq:fly_coupling}, the $\theta_{ij}$ represents the angle between the goal vectors of spin $i$ and spin $j$. The $\eta$ parameter sets the ``shape" that the animal uses to encode its space. Finally the velocity that the animal moves at is shown below in Equation \ref{eq:fly_velocity}. In this equation $\hat{\pi_i}$ represents the goal vector for each individual spin that points from the current position to that specific neurons target. 

\begin{equation}
    \label{eq:fly_velocity}
    \vec{V} = \frac{v_0}{N}\sum \hat{p_i}s_i
\end{equation}

At each step, the system performs "neuron updates" which are encoded as sampling updates within the chip. After a set of samples are taken, and the system settles into a local minima, the system moves forward with velocity $\vec{V}$ updating the goal vectors $\hat{p_i}$, the angles $\theta_{ij}$, and the couplings $J_{ij}$. This update state then serves as the seed for the next set of neuron updates. In this way the system can settle into a local attractor as it continuously samples from the state space after each update until it finally picks a direction to travel in. 

The original neural models calls for the state at the end of each ``neuron update" to serve as the seed state for the next update. However the PASS system does not have memory of previous system states between sampling runs. To combat this, we modify the Hamiltonian from above to include memory of the previous time step as shown in equation \ref{eq:fly_hamiltonian_bias}, where we add a bias term from the previous time step. Note that the value of $s^{t-1}_i$ is a constant representing the value of that particular state in the previous sampling run and is not a variable in this situation.

\begin{equation}
    \label{eq:fly_hamiltonian_bias}
    H(s^t) = \frac{-k}{N} \sum_{i \neq j} J_{ij}s^t_is^t_j + \alpha \sum_i s^{t-1}_i s^{t}_i 
\end{equation}

With these modifications, we find that the PASS system is able to reproduce the geometry of decision making in fly and locust populations. In Figure \ref{fig:fly} we demonstrate how the neural tuning parameter $\eta$ moves the decision making point for the fly, and affects the individual geometry for that fly. On the right side of the figure, we overlay the $\eta=1.0$ system on top of a heatmap generated by placing a fly in a virtual reality environment. We find that we are able to accurately reproduce the bifurcation point in decision making of actual flies. This demonstrates the ability of the PASS chip to model decision making in primitive animal populations, with the possibility of expanded connectivity in the chip to be able to model more complex decisions in animal brains. 

\section*{Time to Solution Scaling}

 From a theoretical perspective, \cite{Hayes2007AGraphs}  has proven Equation \ref{eq:mix} below demonstrating a linear speed increase of mixing time with problem size when comparing discrete time mixing $\tau_{mix}$ with continuous time mixing $\tau^C_{mix}$. From an intuitive perspective this can be understood as the spins being able to update completely independently in parallel, increasing the average update rate by a factor of $n$ where $n$ is the number of neurons. 

\begin{equation}
    \label{eq:mix}
    \tau_{mix} \geq \frac{n \tau^C_{mix}}{6}
\end{equation}

We note that the theoretical basis proven in that work relies on full connectivity of the underlying graph. In this case, if we are using a graph-colored update scheme, then we would expect that the PASS performance scales as the degree of connectivity (in the case of king's move connection, this would be degree 8). As the degree of connectivity scales up, we expect the performance of the PASS chip to increase relative to a similarly constructed synchronous chip. 

To understand the scaling advantages of the the asynchronous system we performed a series of scaling experiments on instances of the MaxCut and Sherrington Kirkpatrick (SK) problems. The problems are taken from \cite{Hamerly2019ExperimentalAnnealer}, with 10 problems for each size from 10 to 150 variables. Each of the 10 problems is run for 100 trials in both a synchronous and asynchronous scheme. The synchronous scheme has an update rate that is equal to the $\lambda_0$ parameter extracted from the hardware neuron, which is the same as the average update rate for a single neuron. The asynchronous scheme is run using the asynchronous sampling scheme denoted in the "Simulation Model" section above with the same $\lambda_0$ update rate for each neuron. 

\begin{figure*}
\begin{centering}
\includegraphics[width=0.88\linewidth]{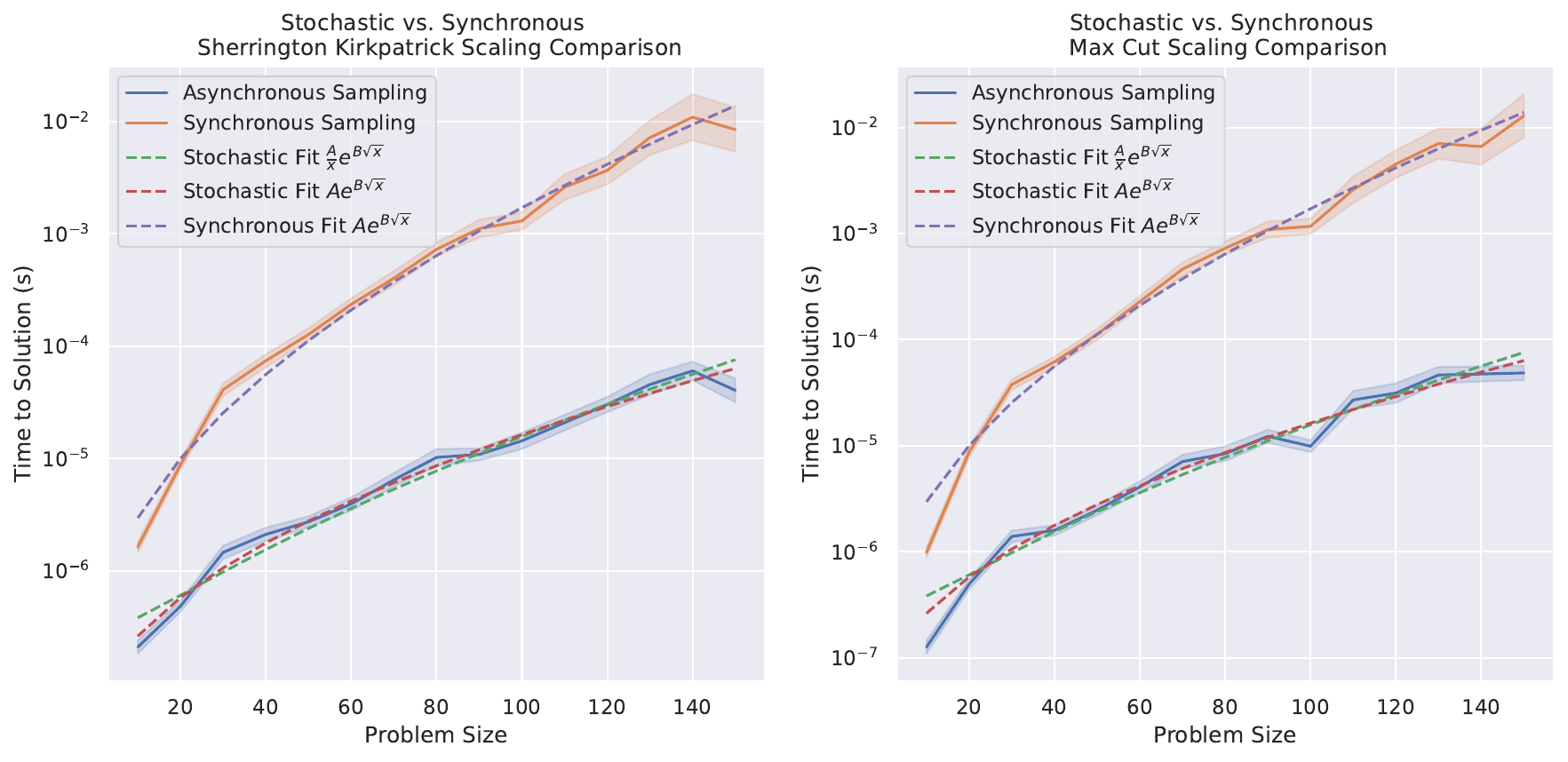}
\par\end{centering}
\caption{\label{fig:SK_MC_scaling_fit}\textbf{Scaling Comparison and Fit for Simulation Model}  \\
The simulation model described in the methods section was used to solve fully connected random MaxCut instances and fully connected random Sherrington Kirkpatrick Instances. The synchronous model is a Gibbs Sampler while the asynchronous model is an asynchronous gibbs sampler. Results show the potential for a large wall clock advantage as well as the potential for a scaling advantage for an asynchronous system compared to an equivalent synchronous one. Shaded areas represent 2 standard deviations around the mean function. }
\end{figure*}

The results of this experiment are shown in Figures \ref{fig:SK_MC_scaling_fit} and Table \ref{tab:scaling_fit_params}. To understand whether the time to solution improvement was linear as theory suggested, we have done a non linear least squares fit on two models for the asynchronous system. The scaling function was chosen to be $Ae^{\sqrt{n}}$ as this is the scaling function that many probabilistic and non-probabilistic systems have exhibited on this same problem set \cite{Hamerly2019ExperimentalAnnealer, Mohseni2022IsingProblems}.  The first model is one where the asynchronous model only provides an improvement that is linear in the problem size, so for a synchronous system that has scaling of $Ae^{B\sqrt{n}}$ we would expect the asynchronous system to have similar exponential fit parameter for $\frac{A}{x}e^{B\sqrt{n}}$ with a potentially different value for $A$. However, when we perform fits over 5000 bootstrapped samples over the trials above, we find that with p<0.01 the exponential fit parameter $B$ is not the same. This suggests that there is a superlinear scaling advantage to moving from a synchronous to an asynchronous system at least demonstrated on the problems shown here. While this is an important result, more work should be done to confirm from a theoretical perspective, as well as a empirical perspective on more problems. 

\section*{Machine Learning Experiments}

To conduct Machine Learning experiments shown here, digits are trained on the PASS system using the modified contrastive divergence algorithm shown in Equation \ref{eq:CD}. The dataset is taken from the MNIST dataset \cite{Lecun1998Gradient-basedRecognition}, and trained on each digit. As the PASS system does not have the necessary neurons to fully represent the data and image, the image is downsampled to the 16 x 16 neuron array, and digit experiments are done 1 digit at a time. This allows the PASS chip to focus on the single digit distributions. We expect that with the addition of further neurons, the PASS system could support more general distributions, and many digits at a time using some neurons as hidden variables \cite{Lecun2015DeepLearning}. 

Training through the contrastive divergence algorithm is done by calculating the expectation of two operators, the data expectation, and the model expectation. The data expectation is calculated in batches of 256 images, and the model expectation is calculated using samples from the PASS system with the given weight value. Training proceeds for multiple epochs over the input data until the distribution converges. To conduct experiments in image reconstruction, we clamp parts of the PASS digit to the input data on programming. This is done using a special clamp bit, which forces a particular neuron output to be 0 or 1. The remaining neurons are sampled from the conditional distribution, given the state of the clamped neurons. 

Estimation of time per sample is done by implementing a Block Gibbs Sampling algorithm in optimized C++ code using the Eigen library for vectorization and mathematical operations. Code is benchmarked on an AMD EPYC 7443P running at 2.8 Ghz, and power is estimated using the AMD uProf profiler while only measuring power output from the cores running the program. To remove idle and I/O power we measure the CPU power consumption before the program is run and subtract this background power from the power consumption for the system. The benchmark is program is run for 10,000,000 samples and power is estimated at each 50\si{ms} interval and averaged over the full sampling run. The time per sample is estimated as the average over this number of samples.


\onecolumn
\clearpage
\section*{Extended Data}
\beginsupplement


    


\begin{table}[ht!]
    \centering
    \begin{tabular}{|c||c|c|}
    \hline System & A (SK) & B (SK) \\ \hline
Asynchronous System ($\frac{A}{x}e^{B\sqrt{x}}$) & 2.074e-07 -  2.673e-07 & 0.864 - 0.895 \\ \hline
Asynchronous System ($Ae^{B\sqrt{x}}$)& 3.413e-08 -  4.376e-08 & 0.588 - 0.620 \\ \hline
Synchronous System ($Ae^{B\sqrt{x}}$) & 1.322-07 - 1.886e-07 & 0.902 - 0.954 \\ \hline

    \hline System & A (MaxCut) & B (MaxCut) \\ \hline
Asynchronous System ($\frac{A}{x}e^{B\sqrt{x}}$) & 1.486e-07 1.877e-07 &  0.898 - 0.927 \\ \hline
Asynchronous System ($Ae^{B\sqrt{x}}$) & 2.443e-08 - 3.083e-08 & 0.622 - 0.650\\ \hline
Synchronous System ($Ae^{B\sqrt{x}}$) & 9.342e-08 - 1.312e-07 & 0.938 -  0.987 \\ \hline

    \end{tabular}
    \caption{ \textbf{Fit parameters for synchronous vs. asynchronous study} \\ These parameters are fitted using 10 problems per size of problem, taken from the dataset provided by \cite{Hamerly2019ExperimentalAnnealer}. Each experiment is run for 100 trials. The ranges shown are 95\% confidence intervals of each parameter generated using 5000 bootstrapped samples within each of the trials. Results show that with p<0.01 confidence we have rejected the hypothesis that the asynchronous and synchronous system have the same exponential fit parameter, suggesting that there is a large scaling advantage to the system.}
    \label{tab:scaling_fit_params}
\end{table}


\begin{figure*}[ht!]
\begin{centering}
\includegraphics[width=0.88\linewidth]{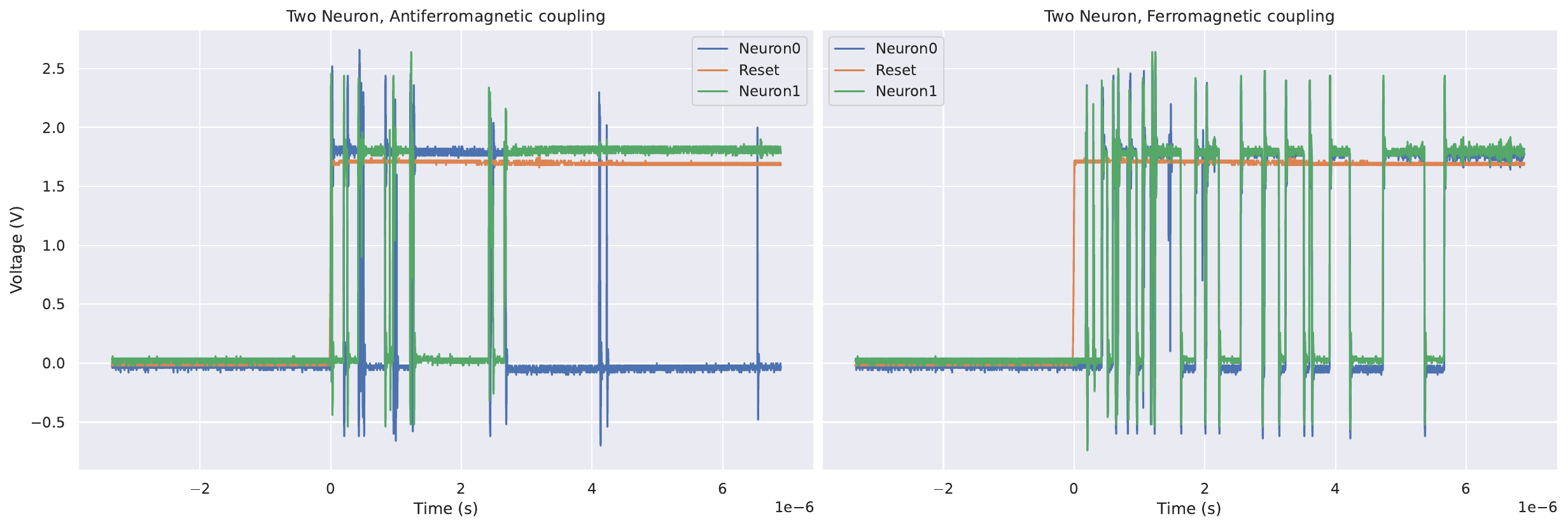}
\par\end{centering}
\caption{\label{fig:AFM_FM}\textbf{Time Domain Analysis of Antiferromagnetic and Ferromagnetic Switching}  \\
\textbf{ Left:} Time domain analysis of two neurons in an antiferromagnetic (negative) coupling.  On reset, the two neurons immediately are pushed into opposite states, and spontaneously switch together based on the input noise. At all switches, we can see that the neurons are in opposite states. 
\textbf{ Right:} Time domain analysis of two neurons in a ferromagnetic (positive) coupling. After reset, the neurons move exactly in tandem. We can see that there is an extremely small delay between a single neuron switch and the other neuron switching with it. }
\end{figure*}


\begin{table}[ht!]
    \centering
    \begin{tabular}{|c|c|c|c|}
         \hline Name & Average Current (\si{uA}) & \% of Total Average Current & Peak Current (\si{uA}) \\ \hline \hline
         Noise Generation & 6.652 & 7.7\% & 6.8 \\ \hline
         Noise Buffer & 19.04 &  22.02\% &  22.232 \\ \hline
         Noise OTA & 25.59 & 29.58\% & 28.916 \\ \hline
         Sigmoid & 26.37 & 30.05\% & 35.821 \\ \hline
         Output Stage &  1.557 & 1.8\% & 1303.03 \\ \hline
         Binary Dot Product &  .3467 & 0.40\% & 51.909 \\ \hline
         C-2C DAC & .2143 &  0.25\% & 4.061 \\ \hline
        Config & 6.712 & 7.76\% & 6.945 \\ \hline
        Total & 86.482 & 100\% & 8031.34 \\ \hline
    \end{tabular}
    \caption{ \textbf{Simulated Single Neuron Power Consumption at Lowest Setting.} \\
    Power consumption of the hardware neuron without switching load. This shows that the neuron consumes approximately 86 \si{uA} with the majority of the power going into the analog components (the analog amplifiers, buffers, and sigmoid). This can be understood by the need for high speed analog components to allow for the system to operate at \si{Mhz} speeds. The analog amplifiers conume high power to support these speeds. The Capacitor DAC (capdac) does not consume power at rest, and only consumes switching power, while the binary dot product only consumes leakage power without switching loads. }
    \label{tab:single_neuron_b0_power}
\end{table}


\begin{table}[ht!]
    \centering
    \begin{tabular}{|c|c|c|}
         \hline Name & Area ($\si{um}^2$) & Percentage of total \\ \hline \hline
        Noise Generation & 4.956 & 0.23\% \\ \hline
        Noise Buffer & 38.44 & 1.82\% \\ \hline
        Noise OTA & 159.375 & 7.55\% \\ \hline
        Sigmoid Generation &  31.096 &  1.47\% \\ \hline
        Binary Dot Product  & 345.50 & 16.36\% \\ \hline
        C-2C DAC  & 976.262 &  46.23\% \\ \hline
        Configuration Circuitry & 188.55 & 8.93\% \\ \hline
        Output inverters & 7.282 & 0.34\% \\ \hline
        Routing/Fill/Empty Space & 360.109 & 17.05\% \\ \hline
        Total & 2111.57 & 100\% \\ \hline
    \end{tabular}
    \caption{\textbf{Single Neuron Area Breakdown} \\
    Area break down of a single neuron. As can be seen from the post layout imagery, the main area consumption is from the C-2C DAC circuitry. This is due to the need for 7-bit DAC circuitry to keep high amounts of precision for the analog neuron. While the C-2C architecture tends to take less area than binary weighted schemes, the high layout parasitics cause the base capacitor size to increase, which thus increase the area of the circuit. We also note that a large portion of area is dedicated simply to fill and routing, which can be reduced by better layout techniques. 
    }
    \label{tab:single_neuron_area}
\end{table}


\begin{table}[!ht]
    \centering
    \begin{tabular}{|c||c|c||c|c|}
	\hline Speed Setting & I @ VDD=0.8V& Power (0.8 V)& I @ VDD=0.6V&  Power (0.6 V)\\ \hline \hline 
    1	& 24mA &    19.2mW &    16mA &  9.6mA \\ \hline
    2	& 26mA & 	20.8mW & 	17mA &	10.2mW \\ \hline
    3	& 29mA &	23.2mW &	19mA &	11.4mW \\ \hline
    4	& 33mA &	26.4mW &	21mA &	12.6mW \\ \hline
    5	& 38mA &	30.4mW &	23mA &	13.8mW \\ \hline
    6	& 49mA &	39.2mW &	28mA &	16.8mW \\ \hline
    7	& 71mA &	56.8mW &	37mA &	22.2mW \\ \hline
    \end{tabular}
    \caption{\textbf{Full Chip Measured Power at Various Speed Settings} \\
    The measured power from the chip core (not including I/O) during computation of a 16x16 MaxCut problem. }
    \label{tab:single_neuron_meas_power}
\end{table}


\end{document}